\documentclass[10pt,twocolumn,letterpaper]{article}
\usepackage[pagenumbers]{cvpr}
\usepackage{amsthm}
\usepackage{tikz}
\usepackage{booktabs}  
\usepackage{xfrac}
\usepackage{array}
\usepackage{cancel}
\definecolor{cvprblue}{rgb}{0.21,0.49,0.74}
\usepackage[pagebackref,breaklinks,colorlinks,allcolors=cvprblue]{hyperref}
\def\paperID{*****}
\def\confName{CVPR}
\def\confYear{2027}
\newsavebox{\huygensimage}
\DeclareCaptionFormat{huygenswrap}{%
  \parshape=0
  \hangindent=-0.66\linewidth
  \hangafter=-11
  #1#2#3\par
}

\newtheorem{theorem}{Theorem}[section]
\newtheorem{proposition}[theorem]{Proposition}

\newtheorem{corollary}[theorem]{Corollary}
\theoremstyle{definition}
\newtheorem{definition}[theorem]{Definition}
\theoremstyle{remark}
\newtheorem{remark}[theorem]{Remark}
\newcolumntype{R}[1]{>{\raggedright\arraybackslash}p{#1}}

\providecommand{\coloneqq}{\mathrel{:=}}
\newcommand{\eqdef}{\coloneqq}
\NewDocumentCommand{\K}{m}{K_{#1}}                           
\NewDocumentCommand{\hK}{m}{\widehat{K}_{#1}}                
\NewDocumentCommand{\Kop}{m}{\mathbf{K}_{#1}}                 
\NewDocumentCommand{\KopN}{m m}{\mathbf{K}_{#1,#2}}           
\NewDocumentCommand{\Gr}{O{s} O{N}}{G_{#1,#2}}                
\NewDocumentCommand{\Gop}{O{s} O{N}}{\mathbf{G}_{#1,#2}}      
\NewDocumentCommand{\Icap}{O{s} O{N}}{I_{#1,#2}}              
\NewDocumentCommand{\Kt}{O{}}{\tilde{K}_{#1}}                 
\NewDocumentCommand{\Set}{m}{\mathcal K_{#1}}

\newcommand{\B}{B}

\newcommand{\corr}{\mathbin{\star}}          
\newcommand{\Nul}{\mathcal N}                 

\title{\vspace{-0.3cm}A Unified Theory of Image Scaling:\\ Point Spread Functions, Discrete Kernels, and Scale-Spaces}

\author{
Assaf Shocher\\
Technion -- Israel Institute of Technology
}

\begin{document}
\vspace{-0.3cm}\maketitle
\vspace{-0.3cm}
\begin{abstract}
We derive the optically aligned system of image-scaling operators $\Kop{s}$ that reproduces optical zoom (exact zoom-out, canonical zoom-in), by identifying a Huygens-like consistency principle in which each pixel acts as a secondary source. It is determined in closed form by the camera's point spread function, itself the operator $\Kop{0}$. The operators compose,
$\Kop{s'}\circ\Kop{s}=\Kop{ss'}$ (band projected for down-up),
and account for both upscaling and downscaling, the two forming a
Moore--Penrose pseudo-inverse pair, $\Kop{\sfrac{1}{s}}$$=$$\Kop{s}^{\dag}$. We show that image scaling, as routinely practiced, is neither optically faithful nor internally consistent; e.g., common anti-aliasing rules generally correspond to no camera and two $\times\sfrac12$ resizes generally do not equal one
$\times\sfrac14$ resize. Gaussian scale-space kernels cannot be exact finite-aperture PSFs. It is an idealized special case of what emerges here: a
generalized scale-space for sampled images, derived entirely from the
camera optics, bidirectional, and not limited to Gaussian kernels. The
system goes beyond scaling and supports composable and invertible sub-pixel geometric transformations. We empirically validate its optical predictions with real captures and noisy optics simulations, and demonstrate its practical impact on the stability of tasks such as super-resolution and classification.
\end{abstract}

\vspace{-0.3cm}\section{Introduction}\label{sec:intro}
Image scaling underlies almost every computer-vision pipeline:
pyramids \cite{burt83}, super-resolution
\cite{irani91,farsiu04,freeman02,dong16}, augmentation, and display.
Choices include bilinear, bicubic \cite{keys81,mitchell88},
Lanczos \cite{duchon79}, and learned resampling \cite{talebi2021}.
Yet common scaling practice is consistently inconsistent.
Composition, commutativity, and round trips generally fail. These failures trace back to an original sin: prescribing resizing rules without requiring them to describe real optics.

Scale-space theory is an elegant answer, first conceived in Japan in
the 1950s \cite{iijima62,weickert99}. From a few axioms, it derives one semigroup of Gaussian kernels \cite{witkin83,koenderink84,lindeberg94}, with discrete
\cite{lindeberg90,lindeberg2024}, affine \cite{lindeberg2011}, and
pyramid \cite{lindebergbretzner03} versions. {SIFT analyses model camera blur and sampling \cite{morel2011,reyotero2016}. Finite-aperture PSFs, however, are necessarily non-Gaussian.}

\begin{figure}[t]
\centering
\sbox{\huygensimage}{%
  \includegraphics[width=\linewidth]
    {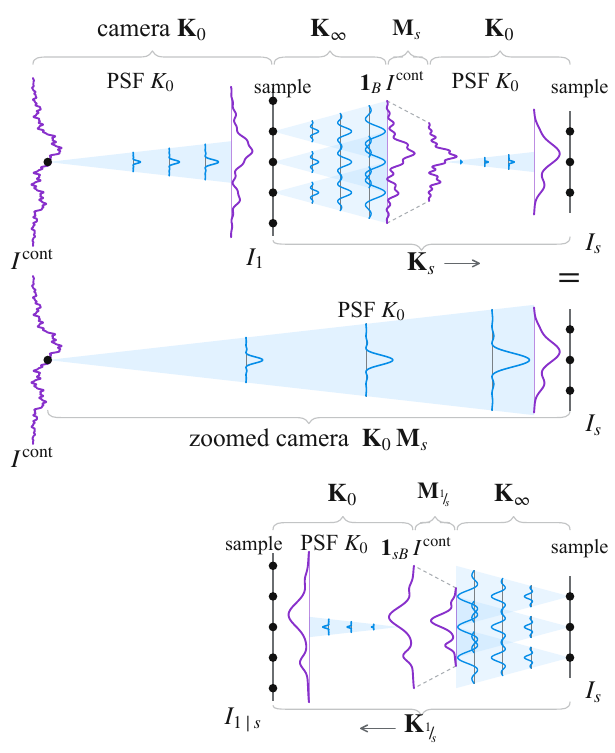}%
}
\usebox{\huygensimage}\par
\vspace{-0.34\ht\huygensimage}

\captionsetup{
  format=huygenswrap,
  font=small,
  textfont=footnotesize,
  singlelinecheck=false,
  justification=justified,
  skip=2pt
}
\caption{A Huygens-like picture of bidirectional scaling. \mbox{$\!s\!<\!1\!$.}
Pixels act as secondary sources whose signals superpose to reconstruct
a band-limited scene. Rescaling and recapturing it with the same camera
(top) reproduces optical zoom-out (middle). The reverse route
(bottom, right to left) restores the original resolution, not discarded
frequencies. Details and derivation at Section~\ref{sec:family}.}
\label{fig:secondary-sources}
\end{figure}

Clues for a unified theory appear across fields. In signal processing,
generalized sampling connects nonideal acquisition to reconstruction
\cite{unser94,eldar2003, huang2006,huang2009} and image-resizing with minimal information loss \cite{unser1995resize,munoz2001,nehab2014}.
 {Precedents include affine-view transfer-function ratios \cite{domke2009}, Gaussian PSF matching \cite{cardoso2015}, astronomical homogenization \cite{boucaud2016}, and zoom-pair PSF recovery \cite{delbracio2012}.} The super-resolution community models degradation as discrete convolution followed by subsampling \cite{efrat2013}. Michaeli\&Irani \cite{michaeli2013} derive the discrete downscaling kernel between integer-related resolutions as a ratio of camera transfer functions, but not cross-scale kernel relations, fractional scaling, or upscaling. KernelGAN's \cite{bellkligler19} $\times4$ kernel derivation from $\times2$ kernel hints at a broader cross-scale structure. The continuous-convolution view \cite{munoz2001,shocher2020cc} connects discrete SR kernels to interpolation. 

Common practice separates two scaling traditions, one consistent but {idealized},
the other trying to imitate photographs but inconsistent across scales. This paper makes these two traditions one: given a camera, we derive
its consistent image-scaling family. Building on generalized sampling,
we show how its PSF determines the kernels across scales and grid
phases, with exact coarsening, canonical refinement, and capture
itself as an endpoint.
Appendix~\ref{app:connections} elaborates on these cross-field connections. Fig.~\ref{fig:secondary-sources} depicts a key finding: for scaling, pixels act as secondary point sources, each contributing
a scale-dependent effective PSF, analogous to Huygens' principle
\cite{Huygens1690Traite} for waves. This operator equivalently deconvolves the sampled image to a continuous band-limited copy of the scene, stretches its coordinates, and recaptures with camera PSF and subsampling. Sampled image scaling becomes regular stretching in a modified vector-space. {Under Theorem~\ref{thm:sampled-converse}'s hypotheses, composition fixes the law; pure AA stretching has no physical PSF (Corollary~\ref{cor:physical-aa}).}

We show that the system (i) \emph{is physically grounded and consistent} by complete derivation and optical simulation (ii) \emph{extends beyond ideal conditions}, by testing real images and noisy signals; and (iii) \emph{matters in applications} such as super-resolution and classification stability.
\section{Preliminaries}\label{sec:prelim}

\subsection{The PSF and image formation}\label{sec:psf}
Photography acts on the continuous scene radiance $I^{\mathrm{cont}}$ by convolution
with the PSF $h\ge0$, $\int h=1$ \cite{goodman}. The sensor samples its pixel
grid to record the image 
\vspace{-0.12cm}\begin{equation}\label{eq:imaging}
I[n] \;=\; (I^{\mathrm{cont}} * h)(n),
\qquad n\in\mathbb{Z}.
\end{equation}
The PSF combines lens blur and pixel integration,
$h=h_{\mathrm{lens}}*h_{\mathrm{pix}}$; the lens term includes
diffraction and aberrations. For an aberration-free circular aperture
it is the Airy pattern
\cite{airy1835}, based on the squared Bessel function
$\big[2J_1(r)/r\big]^2$
\cite{goodman}; other pupils, aberrations, sensors, and media give other
PSFs. Its Fourier transform $\widehat h$ is the optical transfer function
(OTF), whose magnitude is the modulation transfer function (MTF).
We denote its visible-frequency set by
$\B$$\eqdef$$\{$$\omega$$:$$\widehat h(\omega)$$\ne$$0$$\}$ modulo null sets.
We write in one dimension; the construction extends to $d$ dimensions, not assumed separable.
{At a fixed pixel-domain PSF, optical scale changes are modeled
by stretching the scene:} with a stretching operator
$\mathbf M_s(I)=I(\tfrac{\cdot}{s})$, the camera sees
$\mathbf M_sI^{\mathrm{cont}}$; $\mathbf M_{\sfrac{1}{s}}$ returns it
to scene coordinates:
\vspace{-0.12cm}\begin{equation}
   I^{\mathrm{cont}}_s=\mathbf M_{\sfrac{1}{s}}\big(h*\mathbf M_sI^{\mathrm{cont}}\big)
    =\big(s\,\mathbf M_{\sfrac{1}{s}}h\big)*I^{\mathrm{cont}}
    \label{eq:psf}    
\end{equation}

\subsection{Discrete downscaling kernels}\label{sec:mi}
For integer reduction $s$$=$$\sfrac{1}{a}$, convolve the input with a discrete kernel $k$ and subsample:
\vspace{-0.12cm}\begin{equation} 
I_s =(I*k)\!\downarrow_{\sfrac{1}{s}}, \quad s<1.
\end{equation}
Plugging in \eqref{eq:imaging} and \eqref{eq:psf} provides $k$--PSF relation \cite{michaeli2013}:
\vspace{-0.12cm}\begin{equation}\label{eq:mi5}
sh(sx)= \sum_nh(x-n)k[n]
\end{equation}\vspace{-0.35cm}

{
\subsection{Scaling sampled images}\label{sec:scaling}
Typical resizing has four phases (Fig.~\ref{fig:scaling}) \cite{keys81,shocher2020cc}.

\emph{Phase 1, projected grid.} Each output pixel $n$ is projected to its
location $G[n]$ in input coordinates. For $N_0$ input and $N$ output
samples, centered grids give \cite{shocher2020cc,resizeright}
(App.~\ref{app:grid}):
\begin{equation}\label{eq:gridcc}
\Gr[n] \;=\; \tfrac{n}{s}
+ \tfrac{N_0-1}{2} - \tfrac{N-1}{2s}.
\end{equation}
For $N=sN_0$, this reduces to $G[n]=n/s+\tfrac12(1/s-1)$. Sampling is
$\Gop(I^\mathrm{cont})[n]:=I^\mathrm{cont}(\Gr[n])$.

\begin{figure}
\centering
\captionsetup{skip=2pt}
\includegraphics[width=\linewidth,trim={0.25cm 0.26cm 0.25cm 0.27cm},clip]{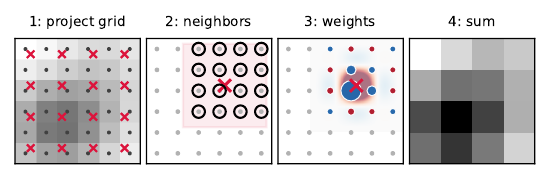}
\caption{\footnotesize The four phases of resizing: output pixels project to sub-pixel locations (crosses), gather neighbors, weight them by chosen interpolation method evaluated at signed offsets (blue positive, red negative) and sum.\vspace{-0.15cm}}
\label{fig:scaling}
\end{figure}

\begin{figure}[b]
\centering
\includegraphics[width=\linewidth]{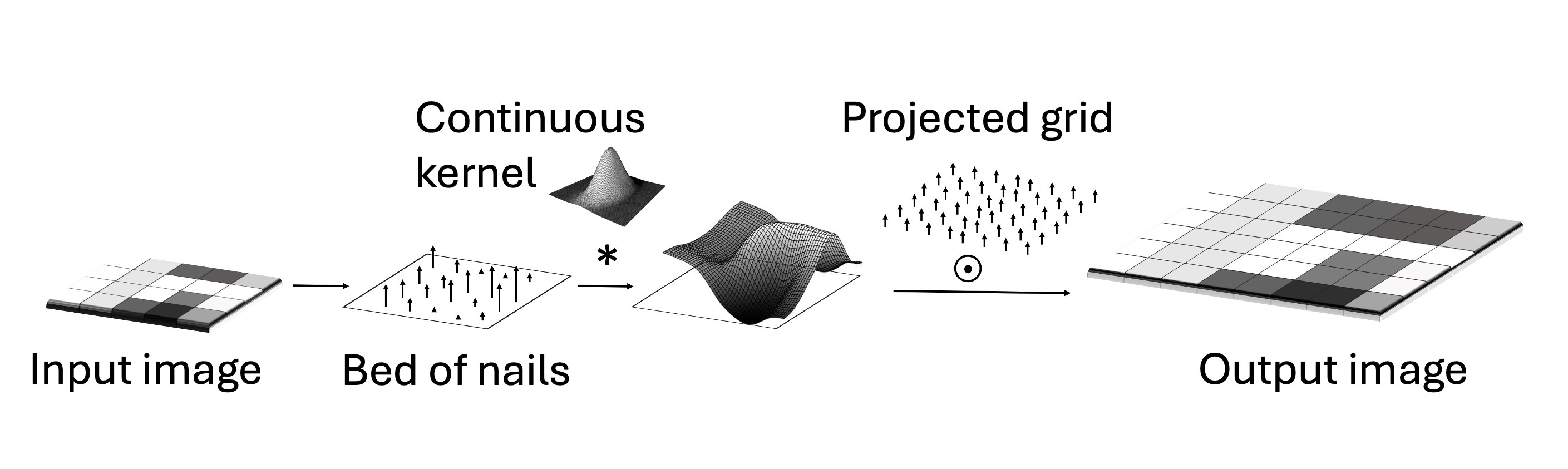}
\caption{\footnotesize{The four phases of Fig.~\ref{fig:scaling} as one continuous process \cite{shocher2020cc}: lifted to a bed of nails, convolve with a continuous kernel, resample.}}
\label{fig:ccprocess}
\end{figure}

\emph{Phase 2, neighbors.} Each $G[n]$ collects the set $\Omega_n$ of input samples inside the kernel support around it.

\emph{Phase 3, weights.} Evaluate the interpolant $\Kt$ at signed offsets: $W[n,m]=\Kt(G[n]-m)$ (Fig.~\ref{fig:prelim}).

\emph{Phase 4, apply.} Each output sample is the weighted sum of its neighbors, $I_s[n]=\sum_{m\in \Omega_n} W[n,m]I[m]$.

\emph{Anti-aliasing.}
Downscaling requires filtering below the output Nyquist frequency to avoid aliasing \cite{shannon49,nyquist28,unser2000}. Common practice approximates this by stretching the interpolant and normalizing its sampled weights:
$
\Kt[s](x)=
\Kt(\tilde{s}x)
\big[\sum_{j\in\mathbb Z}\Kt\big(\tilde{s}(x-j)\big)\big]^{-1},
\quad \tilde{s}=\min(s,1).
$
The denominator is one-periodic, so $\Kt[s]$ remains a single kernel of relative displacement. Thus,
\begin{equation}\label{eq:scaling}
    \Icap[n] = \sum_{m\in \Omega_n} I[m]\;\Kt[s]\big(\Gr[n]-m\big).
\end{equation}
Unless stated otherwise, we assume $N=\lceil sN_0\rceil$.

For constant-reproducing $\Kt$ and integer $a=1/s$,
$\Kt[s](x)=s\Kt(sx)$. When $N=N_0/a$, the centered grid is
$G[n]=an+\tfrac{a-1}{2}$, and \eqref{eq:scaling} becomes
\begin{equation}\label{eq:intkernel}
 I_s=(I*k)\!\downarrow_a,\qquad
 k[m]=s\Kt\!\left(s\!\left(m+\tfrac12(\tfrac1s-1)\right)\right).
\end{equation}
The derivation and rounded-size case are in App.~\ref{app:intkernel}.

\begin{figure*}

\centering
\captionsetup{skip=1.7pt}
\includegraphics[width=\linewidth]{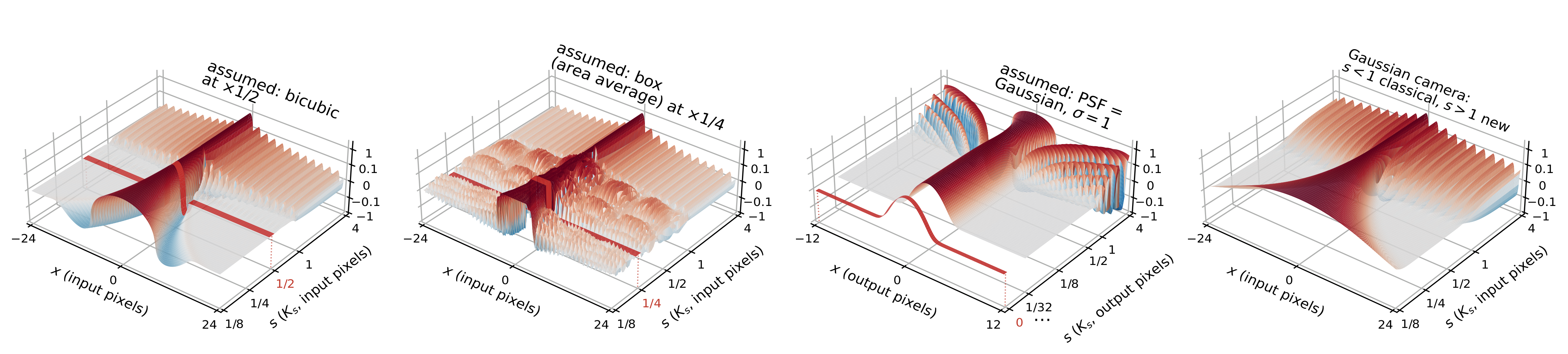}
\caption{\footnotesize Three systems in four views of $\K{s}(x)$ over position and scale (signed-log height). Red: assumed slice (first two) or PSF (third). At $s=1$, $\K{s}$ is a band-limited delta, not a Dirac (Proposition~\ref{prop:endpoints}). In output pixels (third), coarsening approaches the PSF; in input pixels (rightmost), the same Gaussian camera ($\sigma=1$) shows observed-band Gaussian coarsening and band-limited deconvolution for $s>1$ (Remark~\ref{rem:gausscov}). Galleries: App.~\ref{app:figs}.\vspace{-0.25cm}}
\label{fig:surface}
\end{figure*}

\subsection{Image scaling is a continuous convolution}
\label{sec:cc}
Lift the samples to a continuous ``bed of nails,'' $I^\delta(x)=\sum_m I[m]\delta(x-m)$. Convolution with a shifted delta translates the kernel, so
\eqref{eq:scaling} becomes
\vspace{-0.1cm}\begin{equation}\label{eq:ccderiv}
\begin{aligned}
I_s[n] &= \sum_{m} I[m]\big( \delta*\Kt[s] \big)(G[n]-m)\\
       &= \sum_{m} I[m]\int \delta(x-m)\;\Kt[s]\big(G[n]-x\big)\,dx\\
      &= \Gop\big(I^\delta * \Kt[s]\big)[n].
\end{aligned}
\end{equation}
Fig.~\ref{fig:ccprocess}} unifies the four phases of \ref{sec:scaling}; integer reduction $s$$=$$\sfrac{1}{a}$ collapses to a discrete kernel.

\subsection{Scale-space}\label{sec:ss}
Axiomatized in Japan 1959
\cite{iijima62,weickert99} and rediscovered in the West two decades later
\cite{witkin83,koenderink84,lindeberg94}, scale-space theory studies one-parameter families of
progressively smoothed images $\{I * g_t\}_{t\ge0}$. Its founding premise,
that an image exists only through a measurement aperture, is Koenderink's
\cite{koenderink84}, developed as measurement duality by Florack
et~al.~\cite{florack92}. Its axioms are, informally:
(i) \emph{semigroup}, $g_t*g_{t'}=g_{t+t'}$; (ii) \emph{causality}, no new
structure as $t$ grows; (iii) \emph{self-similarity}, all $g_t$ dilations of
one shape. Symmetric, isotropic, self-similar convolution semigroups
of probability kernels admit the stable family
$\widehat g_t(\omega)=e^{-t|\omega|^{\alpha}}$, $0<\alpha\le2$.
Finite variance selects $\alpha=2$, the Gaussian{~\cite{pauwels95,duits04}}. Its
lattice counterpart is the discrete Gaussian
$T(n;t)=e^{-t}I_n(t)$ with $I_n$ the modified Bessel functions
\cite{lindeberg90}.

\subsection{Generalized inverses}\label{sec:pinv}

For a linear operator $A$, a \emph{generalized inverse} $A^{\ddag}$
satisfies $AA^{\ddag}A=A$. The unique one also satisfying
$A^{\ddag}AA^{\ddag}=A^{\ddag}$,
$(AA^{\ddag})^*=AA^{\ddag}$, and
$(A^{\ddag}A)^*=A^{\ddag}A$
is the Moore--Penrose \emph{pseudo-inverse} $A^\dag$; it maps
attainable outputs to minimum-norm preimages
{\cite{moore1920,penrose55,benisrael03}}. For noninvertible $A$, generalized
inverses form an infinite family, $A^{\ddag}=A^\dag+N$ with $ANA=0$.
When $A$ is convolution by $k$, only some are convolutions. Their
Fourier multipliers equal $1/\widehat{k}$ wherever $\widehat{k}\ne0$
and are free at its zeros; choosing zero there yields $k^\dag$.
Thus the convolutional inverses are $k^\ddag=k^\dag+n$, with
$n*k=0$.

\section{Consistent Scaling Operator}\label{sec:family}
We derive a scaling operator by treating pixels as secondary point sources (Fig.~\ref{fig:secondary-sources}), ``emitting'' signals whose superposition reconstructs the band-limited scene, then recaptured by the camera as if no sampling intervened. Analogous to Huygens' wave principle \cite{Huygens1690Traite}: an intermediate image becomes the source of the next, without changing the result.

\begin{definition}[Scaling operators]\label{def:operators}
{For $s>0$, $\Kop{s}$ scales bi-infinite captures in its domain;
its $N$-sample restriction, when a kernel exists, is (in input pixels):}
\begin{equation}\label{eq:opdef}
\KopN{s}{N}(I)=\Gop\big(I^\delta*\K{s}\big).
\end{equation}
The camera operator is
$\Kop{0}(I^{\mathrm{cont}})[n]\eqdef
(I^{\mathrm{cont}}*\K{0})(n)$, where $\K{0}\eqdef h$ is its PSF.
\end{definition}

\begin{definition}[Capture spaces]\label{def:spaces}\quad\\
For $s,r>0$, let $\mathbf 1_{r\B}$ be the scene-space orthogonal
projector onto the frequencies visible at scale $r$. 
\begin{equation}\label{eq:restricted-capture}
I_{s\mid r}\eqdef
\Kop{0}\mathbf M_s\mathbf 1_{r\B}I^{\mathrm{cont}},
\quad
I_s\eqdef I_{s\mid s}
=\Kop{0}\mathbf M_sI^{\mathrm{cont}}.
\end{equation}
The first index is the capture scale; the second is the retained-band
scale and defaults to the first. As $I^{\mathrm{cont}}$ varies, these
captures form $\mathcal V_{s\mid r}$, and
$\mathcal V_s\eqdef\mathcal V_{s\mid s}$ is the full scale-$s$ capture
space. Their spectra are supported in $s\B\cap r\B$.

We call $\B$ dilation-nested if $a\B\subseteq b\B$ whenever
$0<a\le b$; a centered interval has this property. Under
dilation-nested visibility, $I_{s\mid r}=I_s$ whenever $s\le r$:
the retained band then contains everything observable at scale $s$.
\end{definition}

\subsection{The consistency equation and solution set}
\label{sec:derivation}
Assume $\B\subseteq[-\pi,\pi]$ (alias-free); metrics and domains
are in App.~\ref{app:derivation}. Our premises are:
\begin{enumerate}
\item The continuous scene is equivalent to a digital image with
infinitely many pixels: $I_\infty=I^{\mathrm{cont}}$.
\item $\Kop{\infty}\eqdef\Kop{0}^{\dagger}$ is the camera deconvolver.\\
$\Kop{0}\Kop{\infty}=\mathrm{Id}_{\mathcal V_1}$
($\mathcal V_1$ is the range of $\Kop{0}$)\\
$\Kop{\infty}\Kop{0}=\mathbf 1_{\B}$
(projection to the image of $\Kop{\infty}$).
\item Scaling reproduces the part of the matching optical-zoom capture
determined by its input:
$\Kop{s}(I_{1\mid1})=I_{s\mid1}$.
\end{enumerate}
Premise 3 and \eqref{eq:restricted-capture} give
$\Kop{s}\Kop{0}\mathbf 1_{\B}I_\infty
=\Kop{0}\mathbf M_s\mathbf 1_{\B}I_\infty$
for every scene. Right-composing with
$\Kop{\infty}$ and using the Penrose identities gives
\begin{equation}\label{eq:master}
\boxed{\Kop{s}=\Kop{0}\mathbf M_s\Kop{\infty}}.
\end{equation}
Here $\Kop{\infty}$ interpolates and deconvolves a capture
to its canonical camera-visible continuous scene, $\mathbf M_s$
stretches it, and $\Kop{0}$ recaptures it, including sensor sampling.
Equation~\eqref{eq:master} specializes generalized-sampling
rescaling \cite[Eq.~(5.28)]{nehab2014} to canonical reconstruction
\cite{unser94,eldar2003} and same-camera recapture.
 Michaeli\&Irani's discrete downscaling
relation \cite[Sec.~2]{michaeli2013} is recovered at one grid phase
for alias-free integer factors.

\begin{definition}[Consistent kernels]\label{def:set}
$\Set{s}$ is the set of kernels for which \eqref{eq:opdef} realizes the
operator $\Kop{s}:\mathcal V_1\to\mathcal V_s$ of
\eqref{eq:master} on every translated grid, wherever defined.
\end{definition}

{One sampled-grid relation does not yet define a resizer.
Every translated grid requires
$\hK{s}(\omega+2\pi\ell)\hK{0}(\omega)=0$ for $\ell\ne0$,
suppressing noncentral sampling replicas. These constraints give
the solution (App.~\ref{app:derivation}):}
\begin{equation}\label{eq:setfreq}
\hK{s}^{Z}(\omega)=\hK{0}\big(\tfrac{\omega}{s}\big)\,
\hK{0}(\omega)^{\dag}
+Z(\omega)\big(1-\hK{0}(\omega)^{\dag}\hK{0}(\omega)\big),
\end{equation}
$Z$ is free where $\hK{0}$$=$$0$, subject
to \eqref{eq:alias} and operator well-definedness, and irrelevant elsewhere; $\K{s}$ denotes the canonical choice $Z$$=$$0$. If the ratio is unbounded near OTF zeros,
\eqref{eq:setfreq} is {an exact multiplier on its natural domain,
not necessarily an ordinary convolution kernel};
thresholding makes it bounded. Fig.~\ref{fig:surface}
plots the canonical family.

{\begin{corollary}[Pure AA stretching has no physical camera]\label{cor:physical-aa}
Let $h\ge0$ be integrable with $\int h=1$.
No pure-stretch family $\K{s}(x)=s k(sx)$ reproduces its
optical coarsening at both $s=1/a$ and $s=1/a^2$, $a>1$,
even near DC.
\end{corollary}
The obstruction is PSF nonnegativity, not finite kernel support.
Proof: App.~\ref{app:physical-aa}.}

\vspace{-0.1cm}\begin{table}[b]
\centering
\captionsetup{skip=2pt}
\scriptsize
\setlength{\tabcolsep}{1.8pt}\renewcommand{\arraystretch}{1.03}
\setlength{\aboverulesep}{1.5pt}\setlength{\belowrulesep}{1.5pt}
\begin{tabular}{@{}R{0.27\linewidth}!{\color{gray!50}\vrule}R{0.3\linewidth}R{0.2\linewidth}R{0.18\linewidth}@{}}
\toprule
 & kernels & outputs & convolution? \\
\midrule
down, full visible band & \textbf{one} & \textbf{one} & yes \\
\cmidrule[0.2pt](lr){1-4}
down, partial visible band & many, consistent & \textbf{one} & yes \\
\cmidrule[0.2pt](lr){1-4}
up, within the band & many, consistent (one iff full band) & \textbf{one}, band-limited interpolation & yes \\
\cmidrule[0.2pt](lr){1-4}
up, adding frequencies & many, none consistent & \textbf{many}, $\Kop{s}I+\operatorname{null}\Kop{\sfrac{1}{s}}$ & not necessarily \\
\bottomrule
\end{tabular}
\caption{{\footnotesize Uniqueness: closed working band,
dilation-nested visibility. Null spaces permit many kernels
but one output; new frequencies allow many outputs.
``Full band'': $\hK{0}\ne0$ a.e. on $[-\pi,\pi]$.}}
\label{tab:onemany}
\end{table}

\subsection{Realization}\label{sec:normal}
Correlating the kernel form of \eqref{eq:master} with the camera,
$\corr$ denoting cross-correlation, gives the normal equation,
\begin{equation}\label{eq:normal}
\K{s} * \big(\K{0} \corr \K{0}\big)
=
\K{0} \corr s\,\K{0}(s\,\cdot),
\end{equation}
{a Toeplitz system after discretization}. Tikhonov regularization
(App.~\ref{app:wiener}) at level $\eta$ replaces the
pseudo-inverse factor $\hK{0}^{\dag}$ in \eqref{eq:setfreq} by the
Wiener-form damping:
\vspace{-0.1cm}\begin{equation}\label{eq:wiener}
\hK{s}^{(\eta)}(\omega)=
\frac{\hK{0}\big(\tfrac{\omega}{s}\big)\,\overline{\hK{0}(\omega)}}
{\big|\hK{0}(\omega)\big|^{2}+\eta^{2}}\,.
\end{equation}
{Finite $\eta$ controls noise amplified at weak camera frequencies
(App.~\ref{app:noisy-refinement}), but makes structural identities
approximate.} Exact bounded identities instead use
\eqref{eq:setfreq} with the thresholded camera symbol
$\hK{0,\eta}=\hK{0}\mathbf 1_{\{|\hK{0}|\ge\eta\}}$; this is the
family used in the exact-identity figures. {Nesting-dependent claims
also require the modified visible set to remain dilation-nested.} The pairwise Wiener instance
is astronomy's PSF-homogenization kernel
\cite{boucaud2016,bernstein2014}.

\begin{figure}[t]
\centering
\captionsetup{skip=0.4pt}
\includegraphics[width=\linewidth, trim={0.1cm 0.25cm 0.1cm 0.25cm},clip]{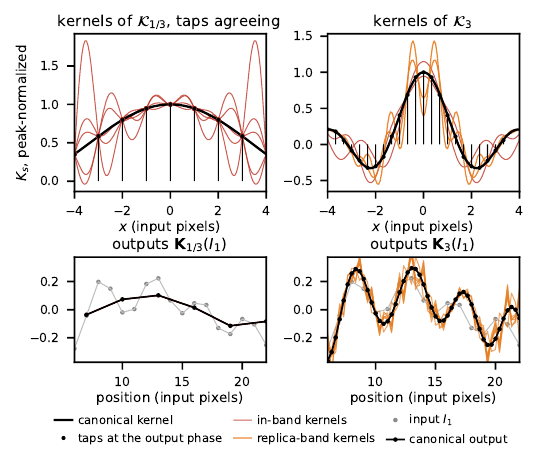}
\caption{\footnotesize Many kernels, one output ($\B=0.6\pi$). Top:
canonical (black), consistent alternatives (red), and replica-reading
nonmembers (orange) at $\times\frac13$ and $\times3$. Members share
taps at $\times\frac13$, but not at
$\times3$. Bottom: consistent kernels give one output; nonmembers add
fine-Nyquist aliases. \vspace{-0.15cm}}
\label{fig:sets1d}
\end{figure}

\begin{figure*}[t]
\centering
\captionsetup{skip=0.2pt}
\includegraphics[width=\linewidth]{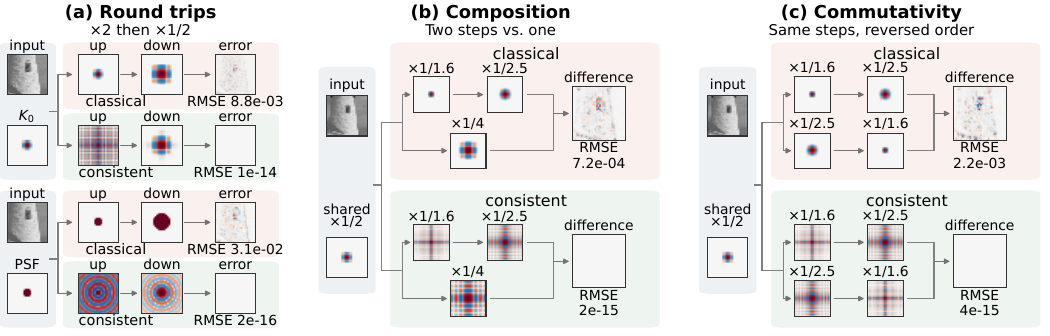}
\caption{\footnotesize
Cycle, composition, and commutativity tests.
(a) Up $\times2$ then down $\times2$ should return the input:
bicubic kernel model (top), defocus-disc camera (bottom).
{The baseline interpolates with the declared kernel and downsamples
with its AA stretch; the disc baseline is constructed, not a library method.}
Ours applies $\Kop{2}$ then $\Kop{1/2}$.
(b) Two unequal, non-integer reductions versus one.
(c) The same reductions in either order.
Both (b,c) use one shared $\times\frac12$ bicubic kernel,
band-limited at $0.99\pi$.
{The baselines fail; ours close to numerical precision.}
More cameras and longer chains: Figs.~\ref{fig:cycleall}, \ref{fig:chain}.}
\label{fig:consistency}
\end{figure*}

\section{Properties of the Operator Family}\label{sec:structure}

\subsection{Composition}\label{sec:comp}
\begin{theorem}[Composition]\label{thm:comp}
For bi-infinite captures in the common domain and compatible
grids, with each operator in its input's pixel coordinates, every
$s$,$s'$$>$$0$ satisfies
\vspace{-0.1cm}\begin{equation}\label{eq:composition-all}
\Kop{s'}\circ\Kop{s}
=\Kop{ss'}\circ
\underbrace{\left(\Kop{\sfrac{1}{s}}\circ\Kop{s}\right)}
_{\mathbf 1_{\B\cap s\B}}.
\end{equation}\vspace{-0.1cm}
\end{theorem}
For dilation-nested $\B$, both sides send $I_1$ to
$I_{ss'\mid\min\{1,s\}}$. Full proof in  App.~\ref{app:compproof}.

\begin{corollary}[Lossless composition]\label{cor:kcomp}
For dilation-nested $\B$, if the chain is not down-then-up
($s$$<$$1$$<$$s'$), then \vspace{-0.1cm}
\begin{quote}\begin{enumerate}
\item $\Kop{s'}\circ\Kop{s}=\Kop{ss'}$.
\item $\K{ss'}=s\,\K{s'}(s\,\cdot)*\K{s}$, when defined.
\item $\Set{s'}\circ\Set{s}\subseteq\Set{ss'}$ under the same condition.
\end{enumerate}\end{quote}
\end{corollary}
Thus the middle cameras cancel unless the first step discarded
frequencies needed by the second.

\begin{corollary}[Inverses]\label{cor:pinv}
Under Theorem~\ref{thm:comp}, with the reverse in its input's
pixel coordinates (App.~\ref{app:compproof}),
\vspace{-0.1cm}\begin{equation}
\Kop{\sfrac{1}{s}}=\Kop{s}^{\dagger}:
\mathcal V_s\longrightarrow\mathcal V_1.
\end{equation}
\end{corollary}

The reciprocal-scale identity gives an arbitrary-scale camera
counterpart to the $L^2$ projection--exact expansion pairing in
nested spline pyramids on compatible grids \cite{unser1993pyramid}.

\begin{corollary}[Commutativity]\label{cor:commute}
For dilation-nested $\B$ and $s,s'\le1$ or $s,s'\ge1$,
$\Kop{s'}\Kop{s}=\Kop{ss'}=\Kop{s}\Kop{s'}$.
Mixed directions obey the projector law \eqref{eq:composition-all}.
\end{corollary}

\begin{theorem}[Sampled composition determines the scale law]\label{thm:sampled-converse}
In one dimension, let real kernels $\K{s}$, $0<s\le1$, have bounded,
compactly supported Fourier multipliers and a coarse limit
$\hK{s}(s\cdot)\to H$ locally in mean square, with $H$ continuous
at zero and $H(0)=1$. If their resizers compose on every
translated grid, then $\B=\{H\ne0\}$ is, up to null sets,
$(-\beta,\beta)$ with $0<\beta\le\pi$, and
\[
\hK{s}(\omega)=H(\omega/s)/H(\omega)\quad\text{on }\B.
\]
Conversely, this ratio, set to zero off $\B$, composes on every
translated grid when bounded.
\end{theorem}
This band-limited converse excludes finite-support interpolants;
its limit need not be physical, and actions outside $\B$ remain free.
Sampling domains and proof: App.~\ref{app:converse}. This couples the scale law to replica nonoverlap,
also required for a principal shift-invariant space to be invariant
under all translations \cite[Cor.~5.6, Rem.~5.8]{aldroubi2010invariance}.

\subsection{Kernels, actions, and completions}\label{sec:collapse}
\begin{proposition}[Uniqueness]\label{prop:collapse}
$\Kop{s}$ is unique.\\
Proof in App.~\ref{app:collapseproof}.
\end{proposition}

Multiple consistent kernels exist in both directions, but different kernels need not mean different outputs. Table~\ref{tab:onemany} summarizes the cases. The
simplest case is Fig.~\ref{fig:sets1d}, left: the kernels differ, yet
they agree at the pixel centers, where the bed of nails sits, so the sums in \eqref{eq:opdef} agree. That is a
{special case, not restricted to integer factors
(App.~\ref{app:collapseproof})}. In general two consistent kernels may also differ at the pixel centers,
but only by a filter whose spectrum lies in the camera's null space
within the digital band and off its sampling replicas, where a capture
has no energy (Fig.~\ref{fig:sets1d}, right). Adding frequencies is a
different matter: it is not necessarily a convolution, and when it is,
the kernel reaches beyond the digital band and reads sampling
replicas. Such a kernel
is not in $\Set{s}$ (Fig.~\ref{fig:set}). For $s>1$ under dilation-nested visibility, the possible completions of
$I_1$ are\vspace{-0.1cm}
\begin{equation}\label{eq:completions}
\{x:\ \Kop{\sfrac{1}{s}}x=I_1\}
=I_{s\mid1}+\operatorname{null}\Kop{\sfrac{1}{s}},
\end{equation}
{the feasible images enforced by back-projection \cite{irani91}.
The canonical minimum-norm completion uses
the matching downscaler's least-squares pseudo-inverse.
A down--up cycle is the closest recoverable-subspace approximation.
Other completions add undetermined frequencies, possibly
non-convolutionally; the canonical one adds none.}

\begin{proposition}[Endpoints and midpoint]\label{prop:endpoints}
The operator limits mean pointwise convergence of their canonical
kernel multipliers in endpoint coordinates:\vspace{-0.1cm}
\begin{quote}\begin{enumerate}
\item $\displaystyle\lim_{s\to0}\Kop{s}=\Kop{0}$.
\item $\Kop{1}=\mathrm{Id}_{\mathcal V_1}$.
\item $\displaystyle\lim_{s\to\infty}\Kop{s}
=\Kop{\infty}$.
\end{enumerate}\end{quote}
\end{proposition}

Proof in App.~\ref{app:endpointsproof}. Intuitively, as $s\to0$, the input samples become infinitely dense, so resizing becomes image formation itself:
convolution with the PSF followed by sampling. At $s$$=$$1$, generally many kernels, not just a
Dirac delta, are the identity on band-limited signals.
Lastly, as $s$$\to$$\infty$, $\Kop{s}$ converges to the camera pseudo-inverse:
Shannon--Nyquist sinc interpolation and PSF deconvolution recover the band-limited scene (Fig.~\ref{fig:endpoints}).

\begin{corollary}[One slice generates the family]\label{cor:telescope}
For nested $\B$ and $0<r<1$, Corollary~\ref{cor:kcomp} and
Proposition~\ref{prop:endpoints} give
\mbox{$
\hK{0}(\omega)=\prod_{j\ge1}\hK{r}\big(r^{j}\omega\big)
$}
Delbracio et al.~\cite{delbracio2012} derived this product to recover a
 PSF from one inter-image zoom kernel; here it follows from the
operator family, so that kernel determines every canonical scale
(Fig.~\ref{fig:cascade}). It is also the refinement-mask infinite
product \cite{mallat89,daubechies92}; it describes a physical camera when the recovered PSF is nonnegative.
\end{corollary}

\section{Generalized Scale-Spaces}\label{sec:scalespace}
With isotropy, regularity, and finite variance, classical scale-space
$\{I*g_t\}_{t\ge0}$ is Gaussian under these axioms \cite{lindeberg94}:\vspace{-0.1cm}
\begin{quote}\begin{enumerate}
\item[(A1)] \emph{linearity};
\item[(A2)] \emph{shift invariance};
\item[(A3)] \emph{semigroup}, $g_t*g_{t'}=g_{t+t'}$;
\item[(A4)] \emph{scale invariance}, all $g_t$ dilations of one shape;
\item[(A5)] \emph{causality}: local extrema are not enhanced.
\end{enumerate}\end{quote}
Camera kernels satisfy (A1)--(A2) in fixed coordinates.
(A3) becomes the conditional scale-product law of
Corollary~\ref{cor:kcomp}; mixed chains retain the intermediate band.
{Consistency does not require self-similarity. We retain the
projected scale-product law without requiring relative kernels to
be dilations of one shape or preserve local extrema, even in coarsening.}
{Resolution refinement here differs from inverse-scale-space
regularization flows \cite{scherzer2001,burger2006}.} We add:\vspace{-0.1cm}
\begin{quote}\begin{enumerate}
\item[(B1)] \emph{exactness on the observed band}:
$\Kop{s}(I_1)=I_{s\mid1}$; under dilation-nested visibility this equals
the full target $I_s$ whenever $s\le1$.
\item[(B2)] \emph{bidirectionality}:
$\Kop{\sfrac{1}{s}}=\Kop{s}^{\dagger}$.
\end{enumerate}\end{quote}
Inverse reconstruction links it to spline scale-spaces
\cite{wanglee1998}.
\begin{definition}[Generalized scale-space]\label{def:gss}
The scale-space of a camera is the family $\{\Kop{s}\}_{s>0}$, satisfying
(A1), coordinate-adjusted (A2)--(A3), and (B1)--(B2).
\end{definition}

\begin{remark}[The Gaussian idealization]\label{rem:gausscov}
For an idealized Gaussian PSF of variance $\kappa_2$, coarsening kernels satisfy (A1)--(A5)
in the continuum; $\K{s}$ is Gaussian for $s<1$,
\vspace{-0.15cm}\begin{equation}\label{eq:gaussfamily}
\K{s}(x)=\tfrac{1}{\sqrt{2\pi\,t(s)}}\,e^{-x^{2}/2t(s)},\;
t(s)=\kappa_2\,(s^{-2}-1)
\end{equation}
Here $t(s)$ is additive for fixed-coordinate smoothing, not resizes.
{{Gaussian variance increments underlie SIFT/ASIFT
\cite{lowe2004,yu2011asift}, scale-equivariant resampling
\cite{worrall2019}, and receptive-field cascades
\cite{lindeberg2026}.}}
Refinement ($s>1$) requires $t(s)<0$: not Gaussian smoothing but
band-limited deconvolution (Fig.~\ref{fig:surface}-right).
 The operators are not a group once coarsening loses information; the
no-cutoff idealization is the exceptional case. \end{remark}

{\begin{figure}
\centering
\begin{minipage}[c]{0.6\linewidth}
\includegraphics[width=\linewidth]{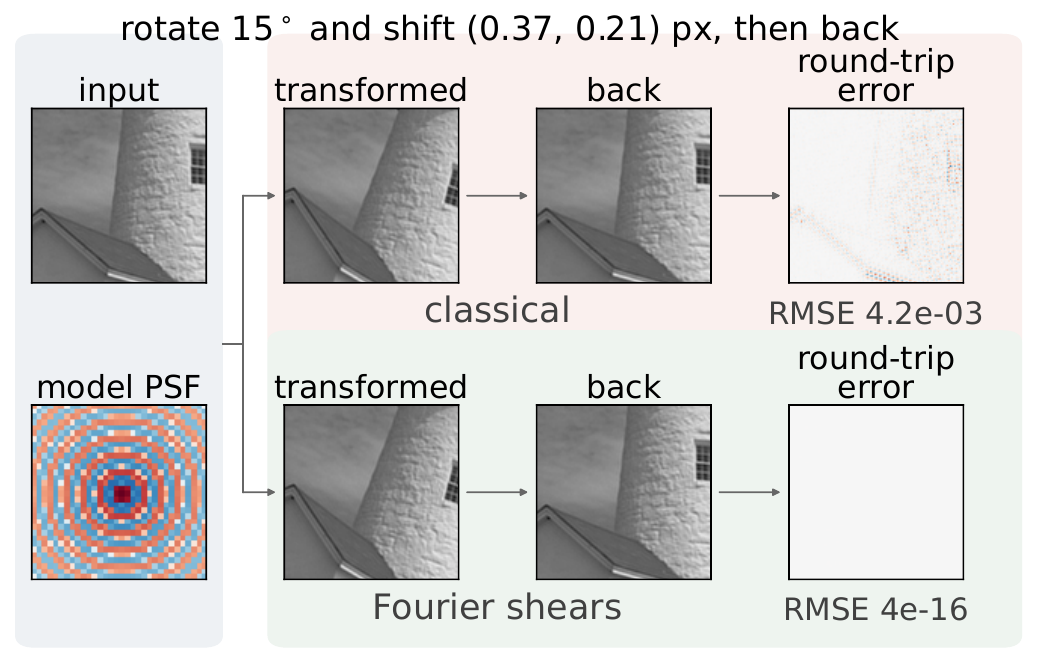}
\end{minipage}\hfill
\begin{minipage}[c]{0.36\linewidth}
\caption{{\footnotesize Geometric transformation test. A capture
is rotated $15^\circ$ and shifted $(0.37,0.21)$ px, then inverted.
Model and details: App.~\ref{app:geometry}.}}
\label{fig:geometry}
\end{minipage}
\end{figure}}

\begin{figure}
\centering
\captionsetup{skip=0.2pt}
\includegraphics[width=0.9\linewidth]{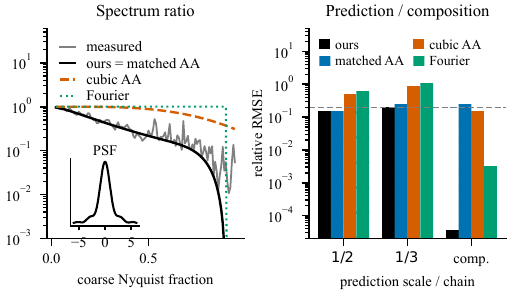}
\caption{\footnotesize Optical test. Left: capture spectrum ratio at $d,2d$.
Right: prediction/composition errors (finite grid); dashed: noise reference.
Matched AA dilates $\K{1/2}$; Fourier is flat on the same band.\vspace{-0.17cm}}
\label{fig:physsim} 
\end{figure}

\vspace{-0.1cm}\section{Beyond scale: camera-coordinate actions}\label{sec:actions}
The construction extends beyond stretching. A linear geometric
action $\mathbf T$ on continuous scenes acts on captures by \vspace{-0.15cm}
\begin{equation}\label{eq:action}
\mathbf A_T\eqdef\Kop{0}\mathbf T\Kop{\infty}.
\end{equation}\vspace{-0.1cm}
This is a camera-conditioned reconstruct--transform--resample
operator \cite{unser1995rotation,munoz2001,huang2009,nehab2014}. Related constructions
use feature-space and Koopman representations
\cite{scholkopf2002,mezic2005}.
If actions bijectively preserve the
camera-visible continuous-scene space, the middle round trip
cancels, so composition and inversion hold in camera coordinates.
Translations preserve every visible set, while a rotation does so when
$\B$ is invariant under that rotation (Fig.~\ref{fig:geometry}). General warps use
\eqref{eq:action} with spatially varying kernels.

\begin{figure*}
\centering
\captionsetup{skip=0.2pt}
\includegraphics[width=\textwidth]{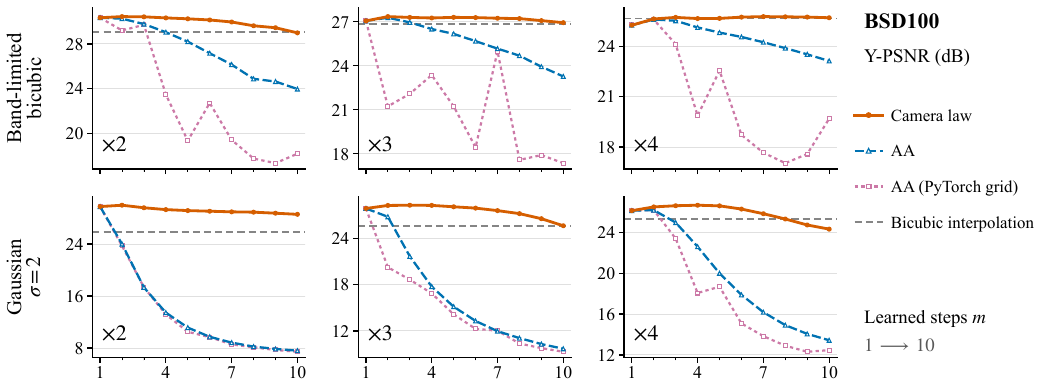}
\caption{{\footnotesize {Scale-law robustness in repeated learned SR:}
BSD100 mean Y-PSNR vs step count.
AA (PyTorch grid) changes only downsampling grid/size;
upsampling remains ResizeRight. Gray: one-shot bicubic.
Vertical ranges differ. Set5/Set14, protocol, and controls:
App.~\ref{app:zssr}.}}
\label{fig:zssr}
\end{figure*}

\section{Conventions Audit and Validation}\label{sec:audit}
\subsection{Inconsistency of common conventions}

{\begin{remark}[AA stretching is not an optical scale law]\label{rem:aa}
Corollary~\ref{cor:physical-aa} excludes pure stretching for every
nonnegative, integrable PSF. Interpolants
reduce to this rule at integer factors (\eqref{eq:intkernel});
sample normalization at other factors and grid effects are audited
separately below. Special cases can still compose:
aligned box averaging does so at integer factors.
Even Gaussian AA uses the wrong variance increment,
$s^{-2}\kappa_2$ rather than $(s^{-2}-1)\kappa_2$
(Remark~\ref{rem:gausscov}).
\end{remark}}

\begin{table}[b]
\vspace{-0.4cm}
\centering
\captionsetup{skip=0.2pt}
\footnotesize
\setlength{\tabcolsep}{4pt}\begin{tabular}{@{}lcccc@{}}
\hline
 & & per-pair fit & consistent & AA stretch \\
\hline
Canon 5D3  & $\times3$ & {32.9} & {31.5} & {31.3} \\
           & $\times4$ & {31.8} & {29.9} & {29.5} \\
Nikon D810 & $\times3$ & {33.0} & {31.6} & {31.0} \\
           & $\times4$ & {32.2} & {30.7} & {30.2} \\
\hline
\end{tabular}
\caption{{\footnotesize RealSR: median PSNR over 50 scenes per
camera. Kernels transfer from $\times2$ with target-pair phase/gain
calibration. Per-pair fits are references, not held-out ceilings
(App.~\ref{app:experiments}).}\vspace{-0.2cm}}
\label{tab:realsr}
\end{table}

\begin{remark}[Up and down are not generally opposites]\label{rem:notopp}
For common finite-support interpolants, up then AA down is {generally} not
the identity (Fig.~\ref{fig:consistency}(a)); {exceptions:
ideal sinc and aligned integer-factor nearest-neighbor up/box down}. One
camera yields the consistent pseudo-inverse pair; an
interpolant/AA-stretch pair generally does not.
\end{remark}

\begin{remark}[{Finite-pupil obstruction}]\label{rem:noguass}
Gaussian and isotropic stable scale-spaces have transfer functions
$\exp(-c|\omega|^\alpha)$ without cutoff; a finite pupil's transfer function,
its autocorrelation, has one. {Thus neither exact family comes from a finite-pupil camera.} A Nyquist-truncated digital approximation gives the
band-limited family of Fig.~\ref{fig:surface}-right; for $\sigma=1$ px,
its omitted Gaussian MTF is only $0.7\%$ of DC at Nyquist, which is why
the Gaussian has served.
\end{remark}

{Cycle, composition, commutativity
(Fig.~\ref{fig:consistency}), long chains (Fig.~\ref{fig:chain}),
and rigid motion (Fig.~\ref{fig:geometry}) test numerical identities,
not camera fidelity. Our declared cameras and ideal sinc close near
machine precision, including radial rigid motion.}

\subsection{\texorpdfstring{{Optical alignment and robustness}}{Optical alignment and robustness}}\label{sec:experiments}\label{sec:optical-tests}

{\emph{A camera computed from optics.}
Simulated planar-texture captures at
$d$, $2d$, and $3d$ include an aberrated pupil, pixel integration,
and read noise; cutoff $0.9\pi$ rad/px \cite{goodman}.
Predicting far captures from the near one gives relative RMSE
$0.15,0.20$, versus cubic AA's $0.53,0.91$ and same-band
Fourier's $0.59,1.05$. AA matched to
$\K{1/2}$ ties at $2d$ but reaches $0.24$ at $3d$, isolating
the scale law. The unregistered power-spectrum ratio follows
$|\hK{1/2}|^2$, not the stretch, until noise dominates
(Fig.~\ref{fig:physsim}). Composition, camera recovery, protocol:
App.~\ref{app:physsim}.}

{\emph{Two DSLRs.} RealSR \cite{cai2019realsr} provides registered
$\times2$, $\times3$, and $\times4$ pairs for 50 scenes each from
Canon 5D3 and Nikon D810 cameras. From each scene's fitted
$\times2$ kernel, we predict $\times3/\times4$ captures by our
camera-derived law or AA stretch. Both use target-pair phase/gain
calibration (Table~\ref{tab:realsr}). Gains are modest; per-pair fits
reach only 32--33 dB amid registration, exposure, and noise.}

{\begin{table}[b]
\vspace{-0.5cm}
\centering
\captionsetup{skip=0.2pt}
\footnotesize
\setlength{\tabcolsep}{4pt}
\begin{tabular}{@{}lrrrrrr@{}}
\toprule
Method & G$\times2$ & G$\times3$ & G$\times4$
& B$\times2$ & B$\times3$ & B$\times4$\\
\midrule
Cubic & 25.80 & 25.59 & 25.27 & 29.67 & 27.22 & 25.98\\
Lanczos & 25.88 & 25.74 & 25.44 & 29.99 & 27.42 & 26.14\\
{Fourier} & {25.87} & {25.78} & {25.50}
& {30.07} & {27.45} & {26.17}\\
{Canonical} & \textbf{30.65} & \textbf{27.62} & \textbf{26.21}
& \textbf{30.10} & \textbf{27.47} & \textbf{26.20}\\
\bottomrule
\end{tabular}
\caption{{\footnotesize One-shot upscaling: mean BSD100 Y-PSNR (dB),
float inputs. G: Gaussian ($\sigma=2$ HR pixels); B: bicubic AA.
{Lanczos-3; ideal Fourier.
Only the canonical operator knows the blur.}
Set5/Set14 and protocol: App.~\ref{app:clean-upscale}.}}
\label{tab:clean-upscale}
\end{table}}

\begin{figure*}
\centering
\captionsetup{skip=0.2pt}
\includegraphics[width=\textwidth]{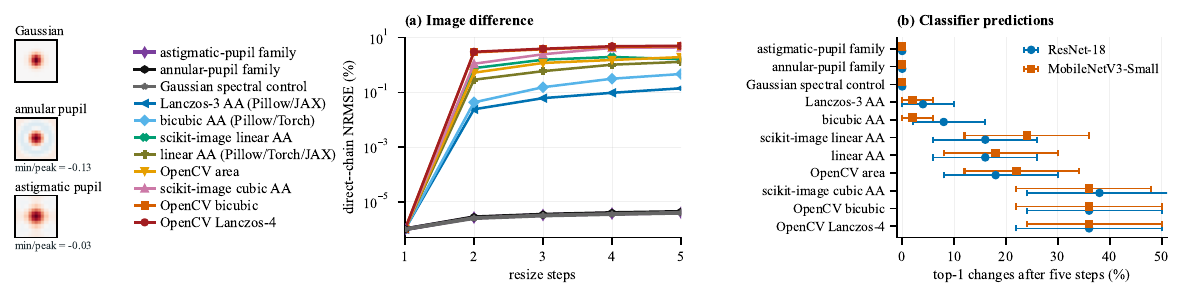}
\caption{{\footnotesize Resize-path dependence on 50 crops,
$2047\to255$ directly or in two to five steps.
Left: signed, peak-normalized $\K{1/2}$, not PSFs; the three
camera families are declared controls, not scene-camera estimates.
(a) NRMSE to each method's direct output. (b) Five-step classifier
prediction changes, with $95\%$ scene-bootstrap intervals.
Protocol: App.~\ref{app:library}; longer chains: Fig.~\ref{fig:chain}.}\vspace{-0.23cm}}
\label{fig:librarypaths}
\end{figure*}

{\emph{Real lenses.} A fixed pixel-domain PSF across zoom
(Section~\ref{sec:psf}) is approximate: aberrations vary.
Center MTF50 \cite{opticallimits} varies about $1\%$
with focal length under diffraction,
but $7$--$11\%$ wide open (App.~\ref{app:lenses}).}

{\emph{Noise, quantization, and aliasing.}
We compare exact and regularized refinement under
noise and quantization; damping controls amplification at weak
camera frequencies (Apps.~\ref{app:noisy-refinement},
\ref{app:clean-upscale}).
An aliasing sweep probes the alias-free model's boundary
(App.~\ref{app:aliasing}).}

\subsection{Effects on applications}\label{sec:applications}

\emph{Upscaling interpolation quality}
Known-blur canonical refinement beats fixed interpolation on clean
BSD100 means (Table~\ref{tab:clean-upscale}); Set5/Set14 and quantization:
App.~\ref{app:clean-upscale}; noise: App.~\ref{app:noisy-refinement}.

\emph{Super-resolution.}
A simplified ZSSR~\cite{shocher18zssr} learns from freshly
downscaled BSD100 crops and reconstructs in
$m$ steps of $S^{1/m}$, $S\in\{2,3,4\}$, $m=1,\ldots,10$.
Only the training downscaler changes; network, budget, and bicubic
upsampling stay fixed; no back-projection. Our law and matched AA
coincide at one step; longer chains test robustness (Fig.~\ref{fig:zssr}),
not optimal SR. Band-limited bicubic gains 2.58--5.02 dB at ten
steps, 0.08--0.17 dB at each arm's best. Set5/Set14 show the same
long-chain trend; protocol and further controls:
App.~\ref{app:zssr-datasets}.

{\begin{figure}
\centering
\captionsetup{skip=0.2pt}
\includegraphics[width=0.95\linewidth,trim=0 10bp 0 0,clip]{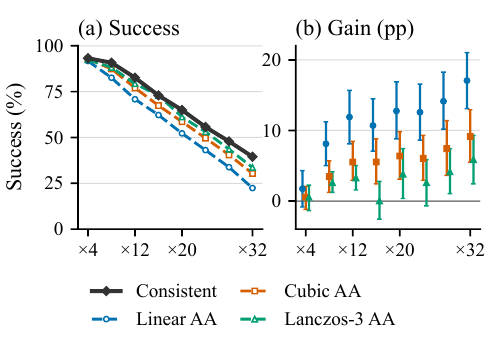}
\caption{{\footnotesize HPatches down--up test. Reduce and restore images before
SIFT homography estimation; success: median error below five
original pixels against supplied truth. Right: consistent-minus-AA
gain with $95\%$ sequence-cluster bootstrap intervals. The consistent
pair realizes ideal retained-band projection for every base.
Protocol: App.~\ref{app:hpatches}.}\vspace{-0.2cm}}
\label{fig:patches}
\end{figure}}

\emph{Libraries and classification.}
Resize choices affect measurements \cite{parmar2022}.
In Fig.~\ref{fig:librarypaths}, library outputs and top-1 predictions
shift across 1--5 steps (50 crops, $2047\!\to\!255$);
our families change negligibly. This is path stability, not accuracy;
matched-RMS noise controls: App.~\ref{app:library}.

{\emph{Down--up feature matching.}
We reduce HPatches images \cite{hpatches_2017_cvpr} by
$4$--$32\times$, restore their size, and score SIFT homographies
against ground truth. At $32\times$, our down--up pair gains
$17.1$, $9.1$, and $5.9$ points over linear, cubic, and Lanczos-3
AA/interpolation pairs; direct low-resolution scores nearly tie
(Fig.~\ref{fig:patches}). Our pair realizes ideal retained-band
projection, not a new interpolator (App.~\ref{app:hpatches}).}

\vspace{-0.2cm}\section{Limitations}\label{sec:limitations}
{Real scale pairs inherit registration resampling; finite tests
truncate kernels, extend boundaries, and exclude margins.
Optical prediction assumes alias-free sampling, not guaranteed by
a finite aperture: aliased inputs may not determine another scale
even when projected operators compose (App.~\ref{app:aliasing}).
We assume one calibrated shift-invariant PSF, though lenses vary with
field and focus; kernel-transfer gains depend on calibration
(App.~\ref{app:psf-misspec}). Regularization trades exact identities for noise control
(App.~\ref{app:noisy-refinement}). We omit
demosaicing, tone curves, and compression.}

\section{Conclusion}\label{sec:conclusion}
Image transformations should inherit the physics of image
formation; we advocate camera consistency as a design criterion
throughout computer vision. Blind SR should estimate a
PSF-represented all-scale family rather than a single kernel.
Optics dictates the observable scale law;
minimum-norm completion fixes canonical refinement.
This scale law was there to be discovered, not invented.


\clearpage
\onecolumn
\appendix
\section{Notation (Sections~\ref{sec:prelim}--\ref{sec:audit})}\label{app:notation}
\begin{center}
\footnotesize 
\begin{tabular}{@{}lp{0.64\linewidth}l@{}}
\toprule
Symbol & Meaning & Defined \\
\midrule
$s$; $r$ & output/input scale; retained-band scale ($s<1$ is coarser) & \S\ref{sec:scaling}, Def.~\ref{def:spaces} \\
$N_0,N$; $n,m$ & input/output sample counts; output/input indices, per axis & \S\ref{sec:scaling} \\
$I^{\mathrm{cont}},I_\infty$ & continuous scene; the same scene viewed as the $s=\infty$ endpoint & \eqref{eq:imaging}, \S\ref{sec:derivation} \\
$I[n],I^\delta$ & a capture; its bed of nails, $\sum_n I[n]\delta(\cdot-n)$ & \eqref{eq:imaging}, \S\ref{sec:cc} \\
$I_{s\mid r},I_s$ & scale-$s$ capture retaining the reference band $r\B$; default $I_s=I_{s\mid s}$ & \eqref{eq:restricted-capture} \\
$\mathcal V_{s\mid r},\mathcal V_s$ & ranges of the preceding captures; $\mathcal V_s=\mathcal V_{s\mid s}$ & Def.~\ref{def:spaces} \\
$h=\K{0}$; $\widehat h=\hK{0}$ & camera PSF; OTF (and MTF $=|\widehat h|$) & \S\ref{sec:psf}, Def.~\ref{def:operators} \\
$\widehat f,f^\vee$ & Fourier and inverse Fourier transforms & \S\ref{sec:psf}, App.~\ref{app:compproof} \\
$\B,s\B$ & camera-visible frequencies; their dilation in reference coordinates & \S\ref{sec:psf} \\
$\mathbf 1_E$ & Fourier projector onto $E$, in the scene or capture coordinates in use & Def.~\ref{def:spaces}, App.~\ref{app:compproof} \\
$\mathbf M_s$ & continuous stretching, $(\mathbf M_sI)(x)=I(x/s)$ & \eqref{eq:psf} \\
$\mathbf T,\mathbf A_T$ & continuous-scene action; its camera-coordinate action & \eqref{eq:action} \\
$\Kop{0},\Kop{\infty}$ & camera; canonical un-capture/deconvolver, $\Kop{0}^{\dagger}$ & Def.~\ref{def:operators}, \S\ref{sec:derivation} \\
$\KopN{s}{N},\Kop{s}$ & scale-$s$ operator restricted to $N$ outputs; its bi-infinite form & Def.~\ref{def:operators} \\
$\K{s},\hK{s}$ & canonical scaling kernel ($Z=0$); its Fourier multiplier & Def.~\ref{def:set}, \eqref{eq:setfreq} \\
$\K{\infty},\hK{\infty}$ & kernel/multiplier of $\Kop{\infty}$, with $\hK{\infty}=\hK{0}^{\dagger}$ & Prop.~\ref{prop:endpoints} \\
$\K{s}^{Z},Z$ & consistent kernel and its free camera-null multiplier & \eqref{eq:setfreq} \\
$\Set{s}$ & all kernels realizing the unique action $\Kop{s}$ & Def.~\ref{def:set} \\
$\dagger,\ddagger;\Nul$ & Moore--Penrose inverse; generalized inverse; camera null space & \S\ref{sec:pinv} \\
$\Gr[n],\Gop$ & centered output grid in input coordinates; sampling on it & \eqref{eq:gridcc} \\
$q$ & first projected coordinate, $q=(N_0-1)/2-(N-1)/(2s)$ & \eqref{eq:gridcenter} \\
$\Omega_n$ & input samples inside the kernel support around $\Gr[n]$ & \S\ref{sec:scaling} \\
$\Kt,\widehat{\Kt},\Kt[s]$ & software interpolant; its transform; normalized AA dilation & \S\ref{sec:scaling} \\
$k,k_{1/a}$ & discrete taps; taps for integer downscale factor $a$ & \S\ref{sec:mi}, App.~\ref{app:intkernel} \\
$\circ,*,\corr,\downarrow_a$ & composition; convolution; correlation; subsampling by $a$ & \S\ref{sec:cc}, \eqref{eq:normal}, \S\ref{sec:mi} \\
$t(s),\kappa_2$ & Gaussian scale-space parameter; PSF variance & Rem.~\ref{rem:gausscov} \\
$\eta$ & Tikhonov level or hard cutoff, as specified & \eqref{eq:wiener} \\
\bottomrule
\end{tabular}
\end{center}

\section{Extended discussion: connections across fields}
\label{app:connections}
Sampling, optics, super-resolution, and scale-space describe resolution
through reconstruction operators, PSF-matching filters, degradation
kernels, and scale families (Section~\ref{sec:intro}). Here we relate
these objects through one camera.

\paragraph{From acquisition to resampling.}
Consistent reconstruction for nonideal acquisition \cite{unser94},
oblique projections and generalized inverses \cite{eldar2003}, and
least-squares resizing with continuous, scale-dependent kernels
\cite{unser1995resize,munoz2001,nehab2014} lead to our master equation:
canonical reconstruction followed by same-camera recapture. We derive
its cross-scale family, projected composition, and reciprocal-scale
pseudo-inverses.

\paragraph{From PSF matching to a scale family.}
Transfer-function ratios relate blurred affine views \cite{domke2009};
Gaussian PSF matching extends to nonlinear resampling
\cite{cardoso2015}; and regularized kernels match astronomical PSFs
\cite{boucaud2016}. These account for input blur. Here one camera PSF
also ties the relative kernels and retained information across scales.

\paragraph{From SR kernels to continuous resizers.}
Michaeli and Irani \cite{michaeli2013} separate the discrete downscaling
kernel from the PSF, derive a sampling-aware least-squares solution,
and estimate it by patch recurrence. KernelGAN \cite{bellkligler19}
estimates such kernels from internal patches and builds $\times4$ from
$\times2$. Continuous convolution \cite{munoz2001,shocher2020cc}
extends grid-specific kernels to resampling. Our family relates them
across real scales and phases, including canonical refinement.

\paragraph{From PSF recovery to a camera endpoint.}
An infinite cascade of dilated zoom kernels recovers a PSF
\cite{delbracio2012}; our telescoping relation gives this construction
(Corollary~\ref{cor:telescope}). Capture and deconvolution are the two
endpoints of the same scaling family.

\paragraph{From observation scales to camera scales.}
Classical scale-space links observation scales
\cite{koenderink84,lindeberg94}; nested spline spaces admit
least-squares reduction and exact expansion \cite{unser1993pyramid},
and generalized sampling connects to B-spline scale-space
\cite{wanglee1998}. Scale-equivariant networks use Gaussian-adjusted
smoothing \cite{worrall2019} or ideal low-pass downsampling
\cite{rahman2023}; generalized Gaussian receptive fields admit
infinitesimal relations and smoothing cascades \cite{lindeberg2026}.
Our scale-space instead follows one possibly non-Gaussian camera PSF:
its visible bands determine the kernels and canonical reverse action
under the stated sampling and visibility conditions.

\section{The projected grid (Section~\ref{sec:scaling})}\label{app:grid}

A resize needs both scale $s$, which sets output spacing $1/s$ in input
coordinates, and integer size $N$, which sets sample count. Conflating
them causes half-pixel shifts and failed size round trips.

For input centers $0,1,\ldots,N_0-1$, the $N$ projected output centers are
\begin{equation}\label{eq:gridq}
g_n=\frac{n}{s}+q,\qquad n=0,\ldots,N-1,
\end{equation}
where $q$ is the origin. Aligning the optical axes sets the projected output center
$q+\tfrac{N-1}{2s}$ equal to the input center $\tfrac{N_0-1}{2}$:
\begin{equation}\label{eq:gridcenter}
q+\frac{N-1}{2s}=\frac{N_0-1}{2}\quad\Longrightarrow\quad
q=\frac{N_0-1}{2}-\frac{N-1}{2s},
\end{equation}
which gives \eqref{eq:gridcc}.

\begin{figure}[h]
\centering
\includegraphics[width=0.94\linewidth]{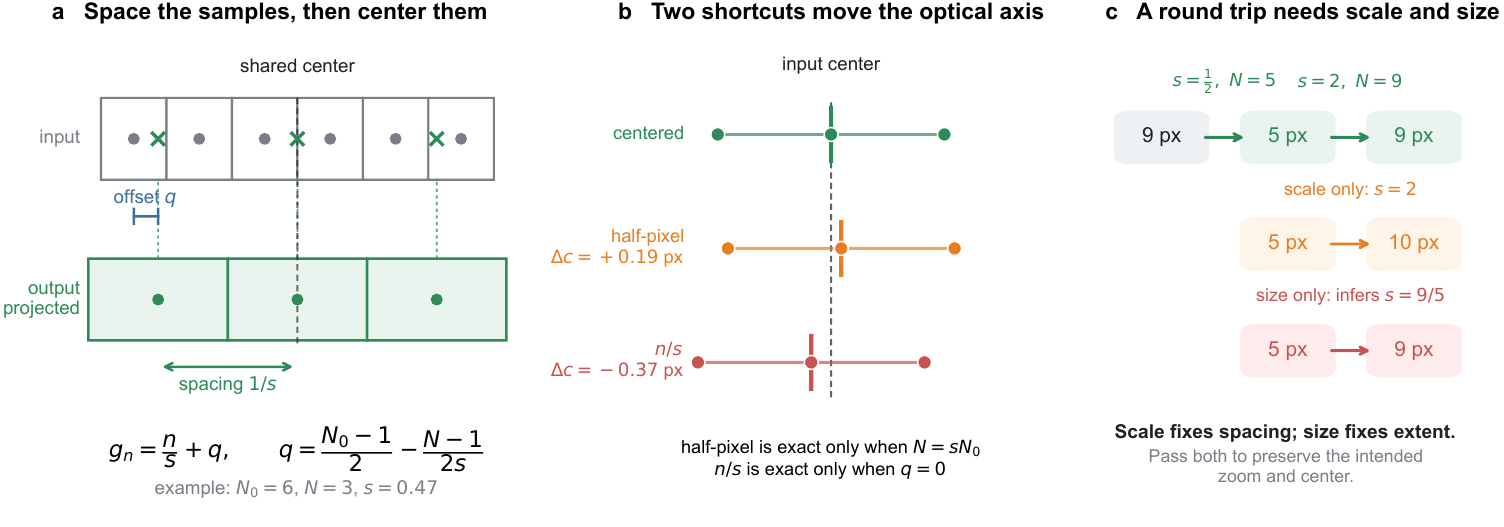}
\caption{\footnotesize The projected grid. (a) Scale fixes spacing; centering
fixes $q$. Crosses show the output centers on the input row. (b) When
$N\ne sN_0$, the half-pixel and first-center shortcuts move the optical
axis. (c) A size-changing round trip needs both its physical scale and
target size.}
\label{fig:grid}
\end{figure}

\emph{Why the shortcuts fail.}
The rule $g_n=n/s$ aligns first pixels, not optical axes. The half-pixel
rule $g_n=n/s+\tfrac12(1/s-1)$ centers only when $N=sN_0$; otherwise it
shifts the output center by
\begin{equation}\label{eq:grid-drift}
\Delta c=\frac{N-sN_0}{2s}.
\end{equation}
Using $N/N_0$ changes the requested zoom. For $N_0=6$, $s=0.47$,
$N=3$ (Figure~\ref{fig:grid}), the centered grid is
$(0.372,2.5,4.628)$; half-pixel and first-center rules shift its
center by $+0.191$ and $-0.372$ pixels.

\emph{Why an inverse needs both numbers.}
Nine pixels at $s=1/2$ produce five samples. Returning at $s=2$
gives ten if size is inferred; requesting nine alone implies scale
$9/5$. The inverse needs both $s=2$ and $N=9$, as supported by
ResizeRight's \texttt{scale\_factors} and \texttt{out\_shape}
\cite{resizeright,shocher2020cc}.

\emph{Why centered grids compose.}
For sizes $N_0\to N_1\to N_2$ and offsets $q_{01},q_{12}$,
\begin{equation}\label{eq:grid-compose}
G_s\!\left(G_{s'}[n]\right)
=\frac{n}{ss'}+q_{01}+\frac{q_{12}}s
=\frac{n}{ss'}+\frac{N_0-1}{2}-\frac{N_2-1}{2ss'},
\end{equation}
the direct centered grid (Theorem~\ref{thm:comp}). Finite code
still needs the stated boundary extension. The formulas apply per axis
with different scales and sizes, without requiring a separable kernel.

\section{Integer-factor reduction (Eq.~\ref{eq:intkernel})}\label{app:intkernel}
Suppose $\Kt$ reproduces constants,
$\sum_{\ell\in\mathbb Z}\Kt(u-\ell)=1$. For $s=\tfrac1a$, $a\in\mathbb N_{>0}$,
\begin{equation*}
\sum_{j\in\mathbb Z}\Kt\!\left(\frac{x-j}{a}\right)
=\sum_{r=0}^{a-1}\sum_{\ell\in\mathbb Z}
 \Kt\!\left(\frac{x-r}{a}-\ell\right)=a.
\end{equation*}
Thus the normalized AA kernel reduces to
$\Kt[1/a](x)=\tfrac1a\Kt(x/a)$. When $N=N_0/a$, the centered grid is
$G[n]=an+\tfrac{a-1}{2}$, and
\begin{equation*}
\begin{aligned}
I_{1/a}[n]
 &=\frac1a\sum_m I[m]\,
   \Kt\!\left(\frac{an+\tfrac{a-1}{2}-m}{a}\right)\\
 &=\frac1a
   \left(I*\Kt\!\left(\frac{\cdot+\tfrac{a-1}{2}}a\right)\right)[an]
  =(I*k)\!\downarrow_a[n],
\end{aligned}
\end{equation*}
with $k$ given by \eqref{eq:intkernel}. The offset
$\tfrac{a-1}{2}$ is the center of the first $a$-pixel block; its
fractional part enters the taps and its integer part enters the padding.
For a rounded output size, the same reduction holds with the general
offset $q$ from \eqref{eq:gridcenter} in place of $\tfrac{a-1}{2}$.

\section{Solving the consistency equation (Section~\ref{sec:derivation})}\label{app:derivation}
Lengths use input pixels; the PSF is $\K{0}=h$, with $H=\hK{0}$.
Assume alias-free sampling, $\B\subseteq[-\pi,\pi]$ up to null sets.
Fourier equalities and band inclusions hold almost everywhere. Scene
space uses the standard $L^2$ inner product; scale-$s$ sequences use
$\langle u,v\rangle_s=s^{-1}\sum_n u[n]\overline{v[n]}$.
The $d$-dimensional density factor is $s^{-d}$. Projections act on
closed capture spaces; unbounded ratios and their compositions use
their natural common domains. An inverse transform satisfying
\eqref{eq:opdef} defines a kernel.

{Explicitly, put $m_s(\omega)=H(\omega/s)H(\omega)^\dagger$.
On the closure of the source capture space its natural domain is
\[
\mathcal D_s=\{I\in\overline{\mathcal V_1}:
m_s\widehat I\in L^2([-\pi,\pi])\}.
\]
Every attainable capture belongs to this domain: if
$\widehat I=H\widehat F$, then
$m_s\widehat I=H(\omega/s)\mathbf 1_{\B}\widehat F\in L^2$.
Thus the operator can be exact without a bounded multiplier or
ordinary inverse-Fourier kernel. Only the spatial derivation below
requires an admissible kernel representative.}

The projector $\Kop{\infty}\Kop{0}=\mathbf 1_{\B}$ gives scene
$I_\infty$ a camera-visible representative $\mathbf 1_{\B}I_\infty$
with the same capture. By
\eqref{eq:imaging}, \eqref{eq:psf}, and
\eqref{eq:restricted-capture},
\begin{equation}\label{eq:twocaps}
\begin{aligned}
I_1[m]&=(I_\infty*\K{0})(m),\\
I_{s\mid1}[n]&=\big((\mathbf 1_{\B}I_\infty)
*s\,\K{0}(s\,\cdot)\big)\big(G[n]\big),
\end{aligned}
\end{equation}
with target samples on the optically centered grid $G[n]$ of
\eqref{eq:gridcc}. Under dilation-nested visibility this is the full
target for $s\le1$; for $s>1$ it is the source-determined part. The resize is
\eqref{eq:opdef},
$\Kop{s}(I_1)[n]=\big(I_1^{\delta}*\K{s}\big)(G[n])$ with
$I_1^{\delta}=\sum_m I_1[m]\,\delta(\cdot-m)$.
Equation~\eqref{eq:master} becomes $\Kop{s}(I_1)=I_{s\mid1}$;
substituting \eqref{eq:twocaps} as target-grid trains gives
\begin{equation}\label{eq:requirefull}
\begin{aligned}
\sum_n \big((\mathbf 1_{\B}I_\infty)*s\,\K{0}(s\,\cdot)\big)(G[n])\,\delta(x-G[n])
=\sum_n \Big[\Big(\sum_m (I_\infty*\K{0})(m)\,\delta(\cdot-m)\Big)*\K{s}\Big](G[n])\,
\delta(x-G[n])
\end{aligned}
\end{equation}
for every $I_\infty$. Write $F=I_\infty$, $K=\K{s}$,
$h_s=s\,h(s\,\cdot)$, $g_s=h_s*(\K{\infty}*h)$, and $x_n=G[n]$.
The continuous target is $F*g_s$.

\paragraph{Spatial reduction.}
Using $\delta(\cdot-m)*K=K(\cdot-m)$ and exchanging the two
sums, the right side of \eqref{eq:requirefull} becomes
\[
\sum_m(F*h)(m)\sum_n K(x_n-m)\,\delta(x-x_n).
\]
Equality of the trains means equality of coefficients. Expanding the
remaining convolution and interchanging sum and integral gives
\[
\begin{aligned}
(F*g_s)(x_n)
 &=\sum_m(F*h)(m)K(x_n-m)\\
 &=\int_{\mathbb R}F(y)\sum_mK(x_n-m)h(m-y)\,dy.
\end{aligned}
\]
Since $F$ is arbitrary, put $u=x_n-y$ to obtain
$g_s(u)=\sum_mK(x_n-m)h(u-(x_n-m))$.
Collect the deltas as the phase-$t$ tap train
$K_t^\delta=\sum_mK(t-m)\delta(\cdot-(t-m))$, we have
\[
K_t^\delta*h=g_s=h_s*(\K{\infty}*h),
\qquad t=x_n\bmod1.
\]
The solution set requires all translated grids, hence all phases
$t\in[0,1)$ rather than one rational grid's phases. At $s=1/a$,
with nested visibility and aligned $t=0$, this is \eqref{eq:mi5}
\cite{michaeli2013}. Averaging over $t$, the intervals $[-m,1-m)$
partition the line and $\int_0^1K_t^\delta\,dt=K$ distributionally.
The continuous kernel equation follows without Fourier:
\begin{equation}\label{eq:masterkernel}
K*h=h_s*(\K{\infty}*h).
\end{equation}
Equivalently, $(K-h_s*\K{\infty})*h=0$: only a camera-null convolution
separates it from the canonical kernel. The all-phase condition is
stronger than its average.

\paragraph{Fourier solution and replica constraints.}
Poisson summation for $K_t^\delta$ transforms the phase equation into
\[
H(\omega)\sum_{\ell\in\mathbb Z}
\widehat K(\omega+2\pi\ell)e^{2\pi i\ell t}
=H(\omega/s)\mathbf 1_{\B}(\omega).
\]
Equating the Fourier-series coefficients in $t$ yields
\begin{align}
\widehat K(\omega)H(\omega)
 &=H(\omega/s)\mathbf 1_{\B}(\omega),\label{eq:pointwise}\\
\widehat K(\omega+2\pi\ell)H(\omega)
 &=0,\qquad \ell\ne0.\label{eq:alias}
\end{align}
Consequently
\begin{equation}\label{eq:solset}
\widehat K(\omega)
=H(\omega/s)H(\omega)^\dagger
 +Z(\omega)\bigl(1-\mathbf 1_{\B}(\omega)\bigr),
\end{equation}
Here $Z$ vanishes on all noncentral replica bands and meets the domain
requirements. This is \eqref{eq:setfreq}; $Z=0$ is canonical. Free
values lie outside \emph{all} source replicas, never on recoverable
new scene frequencies.

For nested visibility and $s\le1$, the projector is redundant; replica
constraints ensure every phase. Refinement needs the source projector
because the input cannot determine missing scene information.

\section{Wiener realization (Section~\ref{sec:normal})}\label{app:wiener}
Let $h_s(x)=s\,h(sx)$, $h\in L^1\cap L^2$, and $\eta>0$.
The Tikhonov realization solves
\begin{equation}\label{eq:wiener-objective}
K_s^{(\eta)}
=\arg\min_{K\in L^2}
\left\{\|h*K-h_s\|_2^2+\eta^2\|K\|_2^2\right\}.
\end{equation}
Replacing $h_s$ by $\mathbf 1_{\B}h_s$ leaves the minimizer unchanged:
the omitted frequencies contribute only a constant to the objective.
The adjoint of convolution by $h$ is convolution by
$h^\sharp(x)=\overline{h(-x)}$; thus $h\corr g=h^\sharp*g$.
Setting the first variation to zero gives
\begin{equation}\label{eq:wiener-normal}
K_s^{(\eta)}*(h\corr h)+\eta^2K_s^{(\eta)}
=h\corr h_s .
\end{equation}
Fourier transformation diagonalizes this equation:
\begin{equation}\label{eq:wiener-solve}
\bigl(|H(\omega)|^2+\eta^2\bigr)\widehat K_s^{(\eta)}(\omega)
=\overline{H(\omega)}H(\omega/s)
\quad\Longrightarrow\quad
\widehat K_s^{(\eta)}(\omega)
=\frac{H(\omega/s)\overline{H(\omega)}}
{|H(\omega)|^2+\eta^2}.
\end{equation}
The positive denominator makes the minimizer unique, proving
\eqref{eq:wiener}. As $\eta\downarrow0$, the multiplier tends pointwise
to $H(\omega/s)H(\omega)^\dagger$; removing the ridge gives
\eqref{eq:normal}. Finite support instead gives a restricted normal
system; the quotient describes unconstrained convolution.

On visible frequencies, $\widehat K_s^{(\eta)}=\widehat K_s q_\eta$ with
$q_\eta(\omega)=|H(\omega)|^2/(|H(\omega)|^2+\eta^2)$.
Regularization attenuates weak camera frequencies, so structural
identities are approximate at finite $\eta$. Exact thresholded
identities use one modified camera
$H_\eta=H\mathbf 1_{\{|H|\ge\eta\}}$ in both numerator and
denominator. Denominator-only clipping does not define that exact
family.

{Coarsening has gain at most one if camera MTF is non-increasing along
each ray: for $s<1$, $|H(\omega/s)|\le |H(\omega)|$. This conditional
bound can fail with interior zeros or nonmonotone defocus or motion
responses. A hard threshold can also create visible-band holes;
nested-band claims must be checked after thresholding.}

{\section{Noisy refinement (Section~\ref{sec:normal})}\label{app:noisy-refinement}
We test realization stability using five $192\times192$ Set5 center
crops, captured at $96\times96$ by an alias-free pupil-derived camera
and enlarged $\times2$. The reference is an independently rendered
optical capture restricted to source-visible frequencies. The separable pupil is
$P(u)=\exp[2\pi i(0.18u^2+0.10u^4)]$ for $|u|\le1$.
Its normalized autocorrelation from 8193 pupil samples defines an OTF
with cutoff $0.9\pi$ per source-pixel coordinate; full-pitch pixel
integration is included. Boundaries are periodic, grids share their
center, and Fourier interpolation splits Nyquist terms before the
fractional grid shift.

We add independent white RGB noise after capture at standard deviations
$\{0,0.25,0.5,1,2,4,8,16\}/255$, with three paired draws per nonzero
level. Exact canonical, Tikhonov, and bicubic refinement share a grid.
The regularizer uses
$\eta=\sigma_n/\operatorname{std}(I)$, with the standard deviation
computed over the noisy RGB input, and preserves observed channel
means. This known-noise scalar rule uses no target pixels and is not
an optimal Wiener estimator. We report whole-image, unclipped BT.601
Y-PSNR with unit peak, averaging dB over draws within images, then images.

Noiseless canonical refinement matches the input-determined target
within $4\times10^{-14}$ per channel. At noise $4/255$, exact,
regularized, and bicubic refinement yield
$-4.03$, $36.72$, and $32.16$ dB, respectively
(Figure~\ref{fig:noisy-refinement}). Weak retained frequencies
amplify noise under exact refinement; regularization controls them
at the cost of exact inverse and composition identities. This tests
one declared camera, not general denoising.

\begin{figure}[t]
\centering
\includegraphics[width=0.4\linewidth]{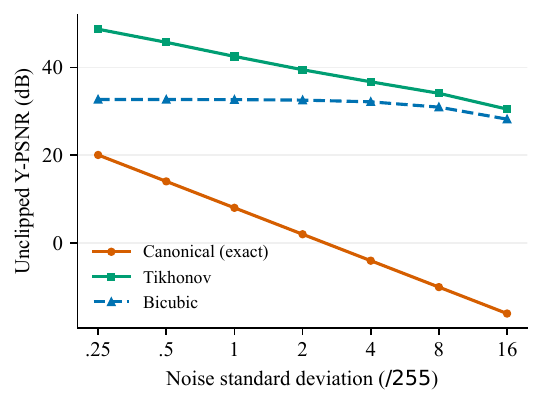}
\caption{{Noise-aware canonical refinement. Mean unclipped Y-PSNR
over five Set5 crops and three noise realizations against the
input-determined optical target. Tikhonov regularization suppresses
noise amplification near weak camera frequencies. No learned model
or missing-frequency prediction is used. The noiseless numerical
identity is reported in the text, not capped on the plot.}}
\label{fig:noisy-refinement}
\end{figure}}

\section{Identifiability from sampled data (Corollary~\ref{cor:telescope})}\label{app:ident}
\begin{remark}[One discrete slice, and when it suffices]\label{rem:oneslice}
Corollary~\ref{cor:telescope} assumes a known canonical continuous
coarsening slice. Recovery from taps without a finite-dimensional
model requires the \emph{complete bi-infinite tap family}.
The offsets are $n/s+q-m$, $n,m\in\mathbb Z$.
For irrational $s$, dense offsets determine a continuous slice.
For reduced rational $s=p/q'$, offsets form a translate of
$p^{-1}\mathbb Z$ and determine a square-integrable slice with centered
Fourier cutoff at most $p\pi$ (up to Nyquist conventions). Without
this band limit, sampling need not be injective. Canonical well-sampled
slices meet the support condition, subject to regularity; noncanonical
free parts need not. Finite taps never determine an arbitrary
continuous slice.
\end{remark}

\begin{remark}[Identifiability from a pair of captures]\label{rem:pair}
An image pair identifies a kernel only if its multirate convolution
map is injective within the chosen class: finite-dimensional designs
need full column rank; ideal infinite alias-free observations can
identify a band-limited kernel by dividing by nonzero source spectrum.
A finite pair cannot identify an arbitrary continuous kernel or camera.
Source zeros, output aliasing, finite extent, and model choice add null
directions beyond camera-null $Z$. Telescoping needs an identifiable
slice and Corollary~\ref{cor:telescope}'s hypotheses.
\end{remark}

\section{Proof of Theorem~\ref{thm:comp} (Composition)}\label{app:compproof}\label{app:k4}
Put $H=\hK{0}$. Indicators denote Fourier projectors in the
scene or capture coordinates in use. The Fourier scaling rule gives
\[
\widehat{\mathbf 1_{\B}\mathbf M_s f}(\omega)
=\mathbf 1_{\B}(\omega)s\widehat f(s\omega)
=s\mathbf 1_{s\B}(s\omega)\widehat f(s\omega)
=\widehat{\mathbf M_s\mathbf 1_{s\B}f}(\omega),
\]
hence
\begin{equation}\label{eq:pf-dilation}
\mathbf 1_{\B}\mathbf M_s=\mathbf M_s\mathbf 1_{s\B}.
\end{equation}
The master equation first gives
\begin{equation}\label{eq:pf-round-op}
\Kop{1/s}\Kop{s}
=\Kop{0}\mathbf 1_{s\B}\Kop{\infty}
=\mathbf 1_{\B\cap s\B},
\end{equation}
the capture-space orthogonal projector in source pixels. Therefore
\begin{equation}\label{eq:pf-many-k}
\begin{aligned}
\Kop{s'}\Kop{s}
 &=\Kop{0}\mathbf M_{s'}\mathbf 1_{\B}\mathbf M_s\Kop{\infty}\\
 &=\Kop{0}\mathbf M_{ss'}\mathbf 1_{s\B}\Kop{\infty}\\
 &=\Kop{ss'}\mathbf 1_{\B\cap s\B}
  =\Kop{ss'}\Kop{1/s}\Kop{s}.
\end{aligned}
\end{equation}
Here $\Kop{\infty}\Kop{0}=\mathbf 1_{\B}$; the last equality
uses the commuting Fourier projectors and
$\mathbf 1_{\B}\Kop{\infty}=\Kop{\infty}$.
This proves \eqref{eq:composition-all} without assuming nested bands.

\paragraph{Lossless composition and commutativity.}
The direct multiplier is supported on $\B\cap ss'\B$, so the
projector is absorbed exactly when
\begin{equation}\label{eq:pf-absorb}
\B\cap ss'\B\subseteq s\B\quad\text{a.e.}
\end{equation}
For nested bands, $s\ge1$ or $s'\le1$ suffices
(Corollary~\ref{cor:kcomp}). If both scales lie on the same
side of one, this holds in either order
(Corollary~\ref{cor:commute}).

\paragraph{Kernel representatives.}
For canonical kernels,
\begin{equation}\label{eq:pf-telescope}
\hK{s'}(\omega/s)\hK{s}(\omega)
=H(\omega/(ss'))H(\omega)^\dagger\mathbf 1_{s\B}(\omega).
\end{equation}
Fourier inversion gives
\begin{equation}\label{eq:pf-kernel-general}
\bigl(s\,\K{s'}(s\,\cdot)\bigr)*\K{s}
=\K{ss'}*(\mathbf 1_{s\B})^\vee.
\end{equation}
For $K\in\Set{s}$ and $L\in\Set{s'}$, the product
$\widehat L(\omega/s)\widehat K(\omega)$ has these same
forced values on $\B$ and inherits the source-replica zeros
of $\widehat K$. Under \eqref{eq:pf-absorb}, it therefore belongs
to $\Set{ss'}$ whenever the convolution is admissible.
This proves the remaining assertions of Corollary~\ref{cor:kcomp}.

\paragraph{Integer-grid realization.}
For strides $a,b\in\mathbb N_{>0}$, substituting the two
convolution sums gives the effective taps
\begin{equation}\label{eq:k4-kill}
k_{1/(ab)}^{\rm chain}
=k_{1/a}*k_{1/b}^{\uparrow a},
\end{equation}
where $k^{\uparrow a}[aj]=k[j]$ and other entries are zero
\cite{bellkligler19}. Grid phases are included in the taps.
This holds for arbitrary discrete kernels; the theorem identifies
the chain with the direct consistent action, not necessarily with
a particular representative's tap array.

\paragraph{Inverses.}
For $A=\Kop{s}$ and $D=\Kop{1/s}$ in their respective input
pixels, $DA=\mathbf 1_{\B\cap s\B}$ and
$AD=\mathbf 1_{\B\cap s^{-1}\B}$.
These project onto the active source and output bands, respectively. Thus
\[
ADA=A,\qquad DAD=D,\qquad (DA)^*=DA,\qquad(AD)^*=AD,
\]
so $D=A^\dagger$. For unbounded exact multipliers, use their
natural domains and the bounded projection extensions of the
round-trip products.

\paragraph{Lost frequencies and finite windows.}
For any visible set, define
\begin{equation}\label{eq:pf-annulus}
A_{s,s'}=(\B\cap ss'\B)\setminus s\B .
\end{equation}
Then, for admissible $I$,
\begin{equation}\label{eq:pf-relax}
\Kop{s'}\Kop{s}I=\Kop{ss'}I
\quad\Longleftrightarrow\quad
\widehat I=0\ \text{a.e.\ on }A_{s,s'}.
\end{equation}
This set has positive width for a nontrivial bounded centered
interval when $s<1<s'$; kernel freedom cannot restore it.
Compatible grids compose by \eqref{eq:grid-compose}, and
\eqref{eq:alias} removes intermediate sampling replicas.
These are full-lattice identities: finite implementations require
a compatible boundary model. The size condition
$N\ge\lceil sN_0\rceil$ alone does not make cropped forward
and reverse matrices a Moore--Penrose pair.

\section{Proof of Proposition~\ref{prop:collapse} (Uniqueness)}\label{app:collapseproof}
Let $A$ satisfy the optical consistency premise. Then
$A\Kop{0}\mathbf 1_{\B}
=\Kop{0}\mathbf M_s\mathbf 1_{\B}$.
For any attainable capture $I$, use
$I=\Kop{0}\Kop{\infty}I$ and
$\mathbf 1_{\B}\Kop{\infty}I=\Kop{\infty}I$ to obtain
\[
AI=A\Kop{0}\Kop{\infty}I
=\Kop{0}\mathbf M_s\Kop{\infty}I=\Kop{s}I.
\]
Thus the action is unique on its common domain, independently
of the definition of $\Set{s}$. Kernel nonuniqueness does not
contradict this: by \eqref{eq:alias}--\eqref{eq:solset},
the difference of any two representatives vanishes on every
source replica, so its convolution with a capture's bed of nails
is zero. \qed

{\paragraph{Equal taps at rational scales.}
For $s=p/q$ in lowest terms and a fixed phase $\delta$, the read
offsets $n/s+\delta-m$ form $\delta+p^{-1}\mathbb Z$.
If the source replicas leave an open spectral gap, choose a nonzero
smooth even $g$ supported away from all $\B+2\pi\ell$, and put
\[
D(x)=[1-\cos(2\pi p(x-\delta))]g^\vee(x).
\]
Its spectrum is a sum of $g$ and its shifts by $\pm2\pi p$,
all in replica-free gaps. Hence $K+D$ remains consistent whenever
$K$ is, while $D$ vanishes at every read offset.
Identical taps therefore do not require an integer factor.
For irrational $s$ the offsets are dense, so two continuous kernels
sharing all these taps must coincide.}

\section{Proof of Proposition~\ref{prop:endpoints} (Endpoints and Midpoint)}\label{app:endpointsproof}
Write $H=\hK{0}$, continuous at zero with $H(0)=1$.
In output pixels the coarsening multiplier is
\[
\widehat{s^{-1}\K{s}(\cdot/s)}(\omega)
=H(\omega)H(s\omega)^\dagger
\xrightarrow[s\to0]{}H(\omega).
\]
In input pixels the refinement multiplier is
\[
\hK{s}(\omega)
=H(\omega/s)H(\omega)^\dagger
\xrightarrow[s\to\infty]{}H(\omega)^\dagger.
\]
At $s=1$, $\hK{1}=HH^\dagger=\mathbf 1_{\B}$, the identity
on attainable captures. These are the pointwise-multiplier
limits stated in the proposition; they do not assert convergence
in operator norm or stability of an unbounded deconvolver.

For completeness, the one-slice corollary follows by finite telescoping:
for $0<r<1$ and dilation-nested visibility,
\[
\prod_{j=1}^{m}\hK{r}(r^j\omega)
=H(\omega)H(r^m\omega)^\dagger
\xrightarrow[m\to\infty]{}H(\omega).
\]
Nesting prevents an intermediate zero from deleting a frequency
where $H(\omega)\ne0$; continuity at zero gives the limit.
Without this condition a spectral hole can obstruct recovery. \qed

\section{Converse from sampled composition
(Theorem~\ref{thm:sampled-converse})}\label{app:converse}
{
The multiplier identity alone does not address sampling. We first
record its consequence, then derive both it and the replica
constraints from composition on every translated grid. All sets and
Fourier identities in the sampled statement are understood up to
Lebesgue-null sets.

\subsection{The multiplier argument (Section~\ref{sec:structure})}
\begin{proposition}[Coarse-limit characterization]\label{prop:converse}
Let finite multiplier representatives $L_s(u)$ satisfy
\begin{equation}\label{eq:converse-law}
L_{st}(u)=L_t(u)L_s(tu),\qquad 0<s,t\le1.
\end{equation}
If $L_s\to H$, with $H$ continuous at zero and $H(0)=1$, then
\begin{equation}\label{eq:converse-ratio}
L_t(u)=H(u)/H(tu)\qquad\text{where }H(tu)\ne0.
\end{equation}
Here convergence may be pointwise, or in $L^2_{\rm loc}$ with each
$L_t$ locally bounded. In the latter case the conclusions hold
almost everywhere. The set $\B=\{H\ne0\}$ is dilation-nested.
Conversely, for a normalized $H$ continuous at zero with nested
$\B$, $L_s(u)=H(u)H(su)^\dagger$ obeys
\eqref{eq:converse-law} and converges pointwise to $H$.
\end{proposition}
\begin{proof}
Let $s\downarrow0$ in \eqref{eq:converse-law}:
$H(u)=L_t(u)H(tu)$. Division gives the ratio, and
$H(u)\ne0\Rightarrow H(tu)\ne0$ gives nesting. In the mean-square
case, dilation preserves local convergence and multiplication by
a locally bounded $L_t$ is continuous on each compact interval.
Conversely,
\[
L_t(u)L_s(tu)
=H(u)\mathbf1_{\B}(tu)H(stu)^\dagger=L_{st}(u),
\]
and continuity at zero gives $H(u)H(su)^\dagger\to H(u)$.
\end{proof}
No bound on the width of $\B$ follows from this multiplier
argument alone.

\subsection{All-phase sampling (Theorem~\ref{thm:sampled-converse})}
Write $F_s=\hK{s}$ and $L_s(u)=F_s(su)$. In the theorem,
$F_s\in L^\infty$ has compact support for each $s$; no common
cutoff, in particular no Nyquist cutoff, is assumed.
Thus $F_s\in L^1\cap L^2$ and $\K{s}$ is a continuous,
band-limited $L^2$ function. Define the infinite-grid resizer
\[
(\mathbf R_s^q I)[n]=\sum_m I[m]\K{s}(n/s+q-m).
\]
The composition hypothesis means, for every finitely supported
input, all $0<s,t\le1$, and every $q_1,q_2\in\mathbb R$,
\begin{equation}\label{eq:sampled-converse-comp}
\mathbf R_t^{q_2}\mathbf R_s^{q_1}
=\mathbf R_{st}^{q_1+q_2/s}.
\end{equation}
There is no intermediate crop or boundary extension.
These sums are well defined: sampling a compact-spectrum $L^2$
function on any fixed lattice gives an $\ell^2$ sequence. Indeed,
periodizing its spectrum produces only finitely many overlapping
replicas, so Parseval gives a finite squared-sample sum. The
sequences depend continuously in $\ell^2$ on the sampling phase.
Cauchy--Schwarz therefore gives absolute convergence and phase
continuity for the products below.

\emph{Composition forces the alias constraints.}
Apply \eqref{eq:sampled-converse-comp} to an impulse at zero.
For any $x$, varying the output phase gives
\[
C(q,x):=\sum_n\K{s}(n/s+q)
\K{t}\big(s(x-n/s-q)\big)=\K{st}(x).
\]
The left side is $1/s$-periodic in $q$. Its Fourier coefficients,
computed by joining the translated integration intervals, are
\[
\begin{aligned}
c_k(x)&=s\int_0^{1/s}C(q,x)e^{2\pi i s kq}\,dq\\
&=s\int_{\mathbb R}\K{s}(y)e^{2\pi i sky}
\K{t}\big(s(x-y)\big)\,dy.
\end{aligned}
\]
Taking the Fourier transform in $x$ yields
$\widehat c_k(\omega)=F_t(\omega/s)F_s(\omega-2\pi sk)$.
Constancy in $q$ is therefore equivalent to
\begin{align}
F_{st}(\omega)&=F_t(\omega/s)F_s(\omega),
\label{eq:sampled-central}\\
0&=F_t(\omega/s)F_s(\omega-2\pi sk),\quad k\ne0.
\label{eq:sampled-replicas}
\end{align}
Conversely, these identities give the constant phase coefficients;
Fourier uniqueness and continuity give the impulse identity for
every phase, hence composition for every finitely supported input.

\emph{The coarse limit determines the ratio and its band.}
Equation~\eqref{eq:sampled-central} is
\eqref{eq:converse-law}. Proposition~\ref{prop:converse} therefore
gives the ratio and dilation nesting. Taking $t=1$ gives
$H=F_1H$, hence $F_1=1$ on $\B$. Taking $t=1$ and $\omega=su$
in \eqref{eq:sampled-replicas}, then $s\downarrow0$ in
$L^2_{\rm loc}$, gives
\[
F_1(u)H(u-2\pi k)=0,\qquad k\ne0.
\]
Consequently $\B\cap(\B+2\pi k)$ is null for $k\ne0$:
the coarse-limit band cannot overlap its sampling replicas.
Real kernels make $H$ Hermitian, so $\B$ is symmetric.
Dilation nesting makes it an interval up to null sets: by Fubini,
for almost every $u\in\B$, almost every point between $0$ and
$u$ also belongs to $\B$. Continuity and $H(0)=1$ ensure a
nonempty neighborhood of zero. Thus
$\B=(-\beta,\beta)$ with $\beta>0$; replica nonoverlap forces
$\beta\le\pi$.

\emph{Converse.}
Let $H$ be a bounded Hermitian function, continuous at zero with
$H(0)=1$, nonzero exactly on $(-\beta,\beta)$ up to null sets,
$0<\beta\le\pi$. Suppose every
$F_s(\omega)=H(\omega/s)H(\omega)^\dagger$ is bounded.
Nesting proves \eqref{eq:sampled-central}. If
$F_t(\omega/s)\ne0$, then $|\omega|<st\beta$; for $k\ne0$,
\[
|\omega-2\pi sk|\ge2\pi s-|\omega|
>s(2\pi-t\beta)\ge s\beta.
\]
This lies outside the support of $F_s$, proving
\eqref{eq:sampled-replicas} and all-phase composition.
Finally $L_s(u)=H(u)H(su)^\dagger\to H(u)$ in
$L^2_{\rm loc}$, since $H(su)\to1$ uniformly on compact
intervals. The boundedness of $H$ in this converse is not an
extra restriction on the forward result: compact $\B$, the ratio
identity, and continuity near zero make the recovered $H$ bounded.

\subsection{What the hypotheses do (Theorem~\ref{thm:sampled-converse})}
\emph{A nondegenerate limit is essential.}
For $\beta=\pi$, let
\[
F_s(\omega)=
\frac{\log(s\beta/|\omega|)}{\log(\beta/|\omega|)}
\mathbf1_{\{0<|\omega|<s\beta\}},\qquad F_s(0)=1.
\]
These bounded real multipliers satisfy both
\eqref{eq:sampled-central} and \eqref{eq:sampled-replicas},
but $F_s(su)\to0$ for almost every $u$. They have no coarse
limit continuous at zero with value one. The normalized limit is
therefore substantive, not merely a choice of units.

\emph{The ratio is determined on $\B$, not on every input space.}
For example, take $\beta=\pi/2$ and
$H(\omega)=(1-|\omega|/\beta)_+$, the transform of a
nonnegative integrable PSF. Besides the canonical family,
\[
F_s(\omega)=H(\omega/s)H(\omega)^\dagger
+\mathbf1_{\{\beta<|\omega|<s\pi\}}
\]
obeys all-phase composition and has the same coarse limit.
The added band disappears as $s\downarrow0$ but permits different
actions outside $\B$. Refinement additionally uses the canonical
pseudo-inverse choice.

\emph{Composition is not optical admissibility.}
Taking $H=\mathbf1_{(-\pi,\pi)}$ satisfies the theorem, but its
inverse transform is the signed sinc, not an intensity PSF.
Nonnegativity of $H^\vee$, and any further optical-pupil
realizability constraints, must be imposed separately.

\subsection{Pure widening (Remark~\ref{rem:aa})}\label{app:pure-widening}
For a pure-stretch family $\K{s}(x)=s k(sx)$, put
$U=\widehat k$. Then $L_s(u)=U(u)$, and
\eqref{eq:converse-law} becomes $U(u)=U(u)U(tu)$.
At $t=1$, $U^2=U$; thus $U=\mathbf1_{\B}$ and $\B$ is
dilation-nested. This conclusion alone does not bound $\B$.
Under the sampled theorem's real-kernel and regularity hypotheses,
all-phase composition further gives
$\B=(-\beta,\beta)$, $\beta\le\pi$. Hence ideal band projection
is the formal consistent pure-widening exception. Its sinc PSF is
signed: it is not an exception to the physical result below.

\subsection{The physical obstruction
(Corollary~\ref{cor:physical-aa})}\label{app:physical-aa}
Let $h\in L^1(\mathbb R)$, $h\ge0$, $\int h=1$, and
$H=\widehat h$. Then $H$ is continuous and $H(0)=1$.
Suppose the pure-stretch family follows optical coarsening for
$s=1/a,1/a^2$, where $a>1$. The camera ratios require, near zero,
\[
H(a\omega)=U(a\omega)H(\omega),\qquad
H(a^2\omega)=U(a^2\omega)H(\omega).
\]
Shrink the neighborhood so the first equality also applies at
$a\omega$ and all three $H$ factors are nonzero. Substituting
then comparing gives $H(a\omega)=H(\omega)$.
For sufficiently small $\omega$, repeated contraction yields
\[
H(\omega)=H(\omega/a^n)\longrightarrow H(0)=1.
\]
This is impossible for a normalized, nonnegative $L^1$ density:
equality $|H(\omega)|=1$ at any nonzero $\omega$ requires
$e^{-i\omega x}$ to be constant on its support almost everywhere.
That support would lie in a countable set, which cannot carry
unit mass for an $L^1$ density. No finite-support or analyticity
assumption on $k$ is needed. The ideal point camera, a Dirac
measure rather than an integrable PSF, is excluded by the stated
hypotheses. The claim concerns a continuous all-phase scale law,
not exceptional composition identities on a single aligned grid.
}

\section{Experimental details: geometric transformation (Figure~\ref{fig:geometry})}\label{app:geometry}
From \href{https://r0k.us/graphics/kodak/kodim19.html}{Kodak image 19},
we take grayscale $[0,1]$ rows $116{:}500$ and columns $64{:}448$
(zero-based, upper endpoint excluded). The $384\times384$ periodic
model has radial symbol
$H(\omega)=\mathbf 1_{\{\|\omega\|<0.65\pi\}}
\prod_{j=0}^{11}\widehat k(\|\omega\|/2^j)$,
for Keys cubic $k$ ($a=-0.5$), whose transform is sampled on
$[-128,128)$ at spacing $1/32$ and linearly interpolated.
This is a formal signed PSF, not a nonnegative optical PSF. After Fourier filtering, we apply the separable
window $w(x)=\tfrac12[1-\operatorname{erf}((|x|-108)/12)]$,
with $x$ measured from the image center $(n-1)/2$; no padding is used.

Both arms rotate by $\theta=15^\circ$, translate by
$(t_x,t_y)=(0.37,0.21)$ px, and use periodic boundaries. Keys
bicubic samples the inverse-coordinate map from $4\times4$ neighbors
modulo $384$. Three-pass Fourier rotation~\cite{unser1995rotation}
uses row, column, row shears with coefficients
$-\tan(\theta/2),\sin\theta,-\tan(\theta/2)$ and one-dimensional
FFT phase shifts; translation uses
$\exp[-i(\omega_x t_x+\omega_y t_y)]$ with $x$ along columns.
The separately specified reverse shifts by $-t$, then shears by
$-\theta$; it is not an inverted matrix.

For a radial camera, rigid motion commutes with convolution and
$\mathbf A_T=\mathbf T$ on the visible space. Finite-grid shear
reversibility does not prove exact Euclidean rotation: windowing leaves
$7.8\times10^{-6}$ of spectral energy outside the declared band,
and real-part extraction at the even-grid Nyquist is not invertible
in general (discarded imaginary values here are below $4\times10^{-16}$).
On the central $160\times160$ crop (rows and columns $112{:}272$),
intensity RMSE is $4.24\times10^{-3}$ for bicubic and
$3.77\times10^{-16}$ for Fourier shears. Dividing by the input
crop's standard deviation $0.1346$ gives $0.0315$ and
$2.80\times10^{-15}$; Figure~\ref{fig:geometry} uses unnormalized RMSE.

\section{Experimental details: library audit (Figure~\ref{fig:librarypaths})}\label{app:library}
From each of the 50 sorted SR-RAW \cite{zhang2019zoom} test-scene folders,
we EXIF-orient \texttt{00001.JPG}, convert to RGB, and center-crop to
$2047\times2047$. Inputs and intermediates are float32 on nominal
$[0,1]$, without quantization or clipping. Square-size paths are
$2047\to255$;
$2047\to723\to255$;
$2047\to1023\to511\to255$;
$2047\to1217\to723\to429\to255$; and
$2047\to1349\to889\to587\to387\to255$.
Scores use the central $224\times224$ output crop (15 pixels removed
at top/left, 16 at bottom/right). Normalized RMSE divides chain--direct
RMSE by the direct crop's pooled-RGB $99.5$th-minus-$0.5$th percentile range.

Pillow 11.3.0 uses per-channel floating-point \texttt{Image.resize}
with bilinear, bicubic, or Lanczos; OpenCV 5.0.0 uses
\texttt{INTER\_AREA}, \texttt{INTER\_CUBIC}, or \texttt{INTER\_LANCZOS4}.
PyTorch 2.8.0 uses bilinear/bicubic \texttt{interpolate} with
\texttt{antialias=True}, \texttt{align\_corners=False}.
JAX 0.4.30 uses \texttt{image.resize}, linear/Lanczos-3, with AA.
Scikit-image 0.24.0 uses \texttt{resize}, orders 1/3, AA,
\texttt{mode=reflect}, \texttt{preserve\_range=True}, \texttt{clip=False}.
Other boundaries use native defaults; output sizes are explicit.
The figure groups close variants under Pillow bilinear, bicubic, or
Lanczos; released tables retain ungrouped measurements.

Controls use centered periodic Fourier synthesis on odd grids, retain
common physical modes, and apply camera-symbol ratios with the center
phase and sampling-density factor.
The Gaussian has $\sigma=0.75$ px. Pupil controls use $401\times401$
samples, cutoff $0.9\pi$, and unit-square pixel integration: the annulus
has inner/outer radius $0.5$, and astigmatism has phase
$2\pi(0.25)(u_x^2-u_y^2)$. Pupil source symbols below $10^{-10}$ are zeroed.
These controls do not estimate the JPEG cameras. {The figure shows
their signed half-scale kernels on a
$13\times13$ input-pixel window, normalized by peak magnitude;
the pupil kernels include negative lobes.}

Torchvision 0.23.0 ResNet-18 and MobileNetV3-Small use
\texttt{IMAGENET1K\_V1} weights in evaluation mode. The scored crop is
clamped to $[0,1]$ and normalized by RGB means $(0.485,0.456,0.406)$
and standard deviations $(0.229,0.224,0.225)$, with no additional resize.
Flip rates compare each chain's top-1 prediction with its direct-resize
prediction. Percentile $95\%$ intervals use 5,000 scene bootstraps;
the plotting script fixes a seed per method/step/model.
Python 3.9.6, NumPy 2.0.2, and Apple's MPS backend were used.

{\paragraph{Different AA policies.}
Pillow, PyTorch AA, and JAX AA widen reconstruction kernels on
reduction; scikit-image uses a scale-dependent Gaussian prefilter.
OpenCV cubic and Lanczos-4 have fixed support, while
\texttt{INTER\_AREA} integrates pixel areas. Camera controls use
transfer-function ratios. Path dependence therefore also reflects
missing AA, grid rules, and boundaries.}

{\paragraph{Matched-size perturbations and multiple crops.}
The supplementary five-step run uses the same 50 scenes and three
overlapping $224\times224$ crops per $255\times255$ output, starting
at $(0,0)$, $(15,15)$, and $(31,31)$. For each method and crop,
we add three independent white Gaussian arrays to the direct output,
each rescaled to the chain--direct RMS. The common reference range
also matches NRMSE. Matching precedes classifier clipping, so
post-clipping norms and spectra may differ. No labels, accuracy
metric, or target-selected noise levels enter.

We average crops and noise draws per scene, then bootstrap 50 scenes
10,000 times (seed 20260921), pairing methods without treating crops
as independent scenes. This run uses Python 3.11.5,
NumPy 1.26.4, Pillow 9.4.0, OpenCV 4.13.0, scikit-image 0.20.0,
PyTorch 2.12.1, and torchvision 0.27.1, with the same V1 model
weights and MPS backend. The archived figure uses the versions above.

\begin{center}\small
\begin{tabular}{lrrlrrl}
\toprule
& \multicolumn{3}{c}{ResNet-18} & \multicolumn{3}{c}{MobileNetV3-Small}\\
Method & Chain & Noise & $\Delta$ 95\% CI & Chain & Noise & $\Delta$ 95\% CI\\
\midrule
Pillow bilinear & 16.7 & 19.6 & $[-11.3,5.1]$ & 16.0 & 37.6 & $[-31.1,-11.6]$ \\
Pillow bicubic & 6.0 & 6.0 & $[-4.7,4.7]$ & 5.3 & 10.4 & $[-10.9,0.4]$ \\
Pillow Lanczos & 2.0 & 2.0 & $[-1.6,1.8]$ & 2.0 & 2.4 & $[-2.4,1.8]$ \\
OpenCV area & 20.7 & 27.8 & $[-16.2,2.0]$ & 27.3 & 36.0 & $[-20.7,2.9]$ \\
OpenCV bicubic & 28.7 & 39.3 & $[-21.8,0.7]$ & 38.7 & 48.9 & $[-20.0,-0.0]$ \\
OpenCV Lanczos-4 & 30.0 & 41.3 & $[-22.0,-0.7]$ & 37.3 & 48.4 & $[-21.1,-0.7]$ \\
scikit-image linear AA & 15.3 & 25.6 & $[-17.8,-3.3]$ & 26.7 & 37.6 & $[-20.2,-1.6]$ \\
scikit-image cubic AA & 34.0 & 49.6 & $[-25.8,-5.3]$ & 44.0 & 55.1 & $[-20.4,-1.8]$ \\
consistent Gaussian & 0.0 & 0.0 & $[0.0,0.0]$ & 0.0 & 0.0 & $[0.0,0.0]$ \\
consistent annular pupil & 0.0 & 0.0 & $[0.0,0.0]$ & 0.0 & 0.0 & $[0.0,0.0]$ \\
consistent astigmatic pupil & 0.0 & 0.0 & $[0.0,0.0]$ & 0.0 & 0.0 & $[0.0,0.0]$ \\
\bottomrule
\end{tabular}
\end{center}
Rates and paired chain-minus-noise intervals are percentage points.
Matched-RMS white noise often causes as many or more flips. The result
shows resize-path dependence, without establishing exceptional
classifier fragility, accuracy gains, or superiority over all
equal-norm perturbations. Camera controls have only numerical path
errors and no flips.}

\section{Experimental details: HPatches (Section~\ref{sec:applications})}\label{app:hpatches}
We use full HPatches sequences, not its pre-extracted patch task
\cite{hpatches_2017_cvpr}: 116
sequences, each with one reference and five targets, yield 580 pairs
with supplied homographies. Both arms use the same grayscale originals.
For the odd periodic Fourier grid, we trim at most three trailing rows
or columns, then test factors $4,8,\ldots,32$ with output dimensions
rounded to the nearest odd integer. The crop origin, and thus the
homographies, stay fixed.

\paragraph{Resize laws.}
The bases are a linear tent, Keys cubic ($\alpha=-0.5$), and Lanczos-3.
Their spatial kernels are cardinal at $1\times$. Treating their
DC-normalized Fourier symbols, truncated to the digital band, as camera
models is an additional assumption; an ordinary Gaussian is not
cardinal and is excluded. Write $\Phi$ for a base symbol and
$r=N_0/N$ for the realized reduction factor on each axis.
Consistent downscaling uses $\Phi(r\omega)/\Phi(\omega)$ on the retained
band; AA stretching uses $\Phi(r\omega)$. Both use the same centered
grid and Fourier modes. The consistent arm restores the original size
with the retained-band pseudo-inverse; AA uses periodic, DC-normalized
unit-width interpolation. Thus the exact consistent down--up map is
the same retained-band projection for all three bases. This tests the
pair, not AA downsampling alone. These are formal kernel-law tests,
not bitwise library runs or physical PSF fits: cubic and Lanczos-3
have signed lobes.

\paragraph{Matching and score.}
Images remain floating point until SIFT input, when they are clipped
to $[0,1]$, multiplied by $255$, and rounded to 8-bit.
SIFT uses up to 1,500 features and contrast threshold $0.01$;
matching uses mutual Lowe-ratio tests at $0.8$.
Pairs with fewer than eight matches fail. RANSAC uses a five-pixel
threshold, 2,000 maximum iterations, confidence $0.995$, and seed $0$.
We compare estimated and supplied homographies on a $5\times5$ grid
spanning $10$--$90\%$ of the reference; points mapped outside the
target by ground truth are omitted.
Success means median disagreement below five original pixels.
The 95\% intervals use 4,000 paired sequence-cluster bootstrap draws
(seed $1337$).
A direct low-resolution control instead uses a 2.5-output-pixel
success threshold and 1.25-output-pixel RANSAC threshold; at
$32\times$ its paired intervals include zero for every base.

{\paragraph{Ideal-projection control.}
On retained modes, the consistent pair cancels its camera multipliers;
its reference is direct ideal Fourier low-pass projection for all
three bases. Across all 116 references at factors $4,16,32$ (1,044
comparisons), float32 outputs differ from direct projection by at most
$2.76\times10^{-7}$ per pixel. Final SIFT rounding changes
$5.52\times10^{-6}$ of values, so bitwise SIFT/RANSAC equality is not
asserted. The gain reflects retained-band preservation by a matched
pair, not improvement over ideal sinc or a distinct kernel-dependent
down--up advantage.}

\section{Experimental details: RealSR (Table~\ref{tab:realsr})}\label{app:experiments}
We use native-size low-resolution images from RealSR v3
\cite{cai2019realsr}. For each scene and $a\in\{2,3,4\}$,
ridge regression fits a $(2R+1)^2$ kernel, $R=7,9,11$, on
{the upper of two disjoint, left-aligned $h\times w$ low-resolution
blocks using all three channels. For image size $H\times W$, set
$m=R+2$, $h=\min(120,\lfloor(H-3m)/2\rfloor)$, and
$w=\min(180,W-2m)$; the zero-based block origins (row, column) are
$(m,m)$ and $(H-m-h,m)$. The ridge weight is chosen from
$\{0.1,1\}$ on the lower block.}
The lower block scores predictions and selects the ridge weight;
it is disjoint from fitting but is not an untouched test set.

The $\times2$ fit is normalized to unit mass and recentered on its
centroid. Its transfer function is evaluated on a $257\times257$
grid over $[-\pi,\pi]^2$, and eleven telescope factors estimate
$\hK{0}$. At $a=3,4$, the consistent multiplier is
$\hK{0}(a\omega)/\hK{0}(\omega)$, restricted to
$|a\omega_i|\le\pi$ on both axes and $|\hK{0}(\omega)|>10^{-2}$.
The stretch evaluates the half-scale kernel's transfer function at
$a\omega/2$. Both predictions use the target pair's fitted centroid
and kernel sum for phase and gain. The comparison is calibrated, not
blind prediction from $\times2$ alone. The fitted reference is limited
to the chosen finite-support model and is not a guaranteed held-out
ceiling. Scores are $\mathrm{PSNR}=-20\log_{10}(\mathrm{RMSE})$ on $[0,1]$.

\section{Experimental details: real lenses (Section~\ref{sec:experiments})}\label{app:lenses}
Real lenses only approximate the fixed pixel-coordinate PSF of
Section~\ref{sec:psf}. Published slanted-edge measurements
\cite{opticallimits} give a scalar check. At f/8, center MTF50 for
the Nikkor Z 24--70 f/2.8 S is
4535, 4574, 4562, 4548 LW/PH at 24, 35, 50, 70\,mm; for the
Sony FE 24--70 f/2.8 GM II it is 4580, 4601, 4623 LW/PH at
24, 40, 70\,mm. The range divided by the maximum is $0.9\%$ in
each case; the Nikkor's f/11 spread is $1.2\%$.
LW/PH means line widths per picture height, not cycles per pixel.
Fixed sensor dimensions within each test preserve the relative
variation. This supports approximate stopped-down center-sharpness
stability, not equal OTFs.
Wide-open center variation is $7$--$11\%$, with greater variation
in the corners, illustrating Section~\ref{sec:limitations}.

\section{Experimental details: the computed camera (Figure~\ref{fig:physsim})}\label{app:physsim}
This one-dimensional simulation derives its camera from a pupil
\cite{goodman}, without fitting a resizing kernel.

\paragraph{Optics and sensor.}
The pupil is
\begin{equation}\label{eq:pupil}
P(u)=\exp\{2\pi i(W_{20}u^2+W_{40}u^4)\}\,\mathbf 1_{[-1,1]}(u),
\qquad W_{20}=0.18,\quad W_{40}=0.10\ \text{waves},
\end{equation}
with defocus and spherical aberration. Its incoherent OTF is
\begin{equation}\label{eq:otf}
H_{\rm lens}(\omega)=
\frac{\int P(u)\overline{P(u-2\omega/\omega_c)}\,du}
     {\int|P(u)|^2\,du},
\qquad \omega_c=0.9\pi\ \text{rad/px}.
\end{equation}
The even pupil gives a real OTF, zero beyond $\omega_c$. We use 4,001
quadrature points and linearly interpolate a 600-frequency lookup on
$[0,1.2\pi]$.
Full-pitch pixel integration gives
\begin{equation}\label{eq:camera}
\hK{0}(\omega)=H_{\rm lens}(\omega)\operatorname{sinc}(\omega/2),
\qquad \operatorname{sinc}(z)=\sin(z)/z .
\end{equation}
The cutoff is below sensor Nyquist, and camera MTF is at least $0.18$
on $|\omega|\le0.6\pi$. Independent read noise has standard deviation
$\sigma=0.005$ on $[0,1]$; shot noise is omitted.

\paragraph{Scene, captures, and operators.}
The scene has $2^{15}$ samples spaced $1/24$ px: Gaussian white noise
(seed $11$), shaped to $1/|\omega|$ amplitude above $1.5$ rad/px,
then set to mean $0.5$ and standard deviation $0.12$.
For $s=1,\tfrac12,\tfrac13$, periodic FFT convolution on this fine
grid produces
\begin{equation}\label{eq:capture}
I_s[n]=
\big(I^{\rm cont}*s\K{0}(s\,\cdot)\big)(n/s)+\varepsilon_n,
\qquad \varepsilon_n\sim\mathcal N(0,\sigma^2),
\end{equation}
sampled at spacing $1/s$ and zero phase at the scene center. Operators
use \eqref{eq:setfreq} on the source digital band with denominator cutoff
$|\hK{0}|>10^{-3}$; the unthresholded numerator makes these numerical
approximations, not the exact thresholded family of
Appendix~\ref{app:wiener}.
One comparator is a unit-mass AA stretch of Keys cubic \cite{keys81}.
The other uses the same camera-derived anchor at $s_0=1/2$:
\[
A_s(x)=\frac{s}{s_0}\K{s_0}\!\left(\frac{s}{s_0}x\right),
\qquad
\widehat A_s(\omega)=\hK{s_0}\!\left(\frac{s_0}{s}\omega\right).
\]
Its anchor multiplier and output equal the consistent arm's; elsewhere
only the scale law changes. The same camera lookup and cutoff define
the anchor, without spatial truncation. A fourth, ideal Fourier arm
has flat transfer on the same visible band:
\[
\widehat F_s(\omega)=\mathbf1_{\{|\omega|<s\omega_c\}},\qquad 0<s\le1.
\]
It retains the output band without correcting the camera's nonflat
response. All four arms use bed-of-nails convolution, identical phases
and target sampling, and independently generated captures. No software
resizer is called. At $1/2$, the matched-AA spectrum-ratio curve
coincides with the consistent curve.

\paragraph{Tests and results.}
Relative RMSE is
$\|\mathrm{prediction}-\mathrm{truth}\|_2/
 \|\mathrm{truth}-\overline{\mathrm{truth}}\|_2$,
excluding 200 samples at each end.
Prediction errors at $\tfrac12,\tfrac13$ are respectively
$0.152,0.200$ for the consistent family, $0.525,0.906$ for cubic
AA, $0.152,0.235$ for matched AA, and $0.592,1.045$ for Fourier.
The shared retained band alone does not reproduce this camera. At
off-anchor $1/3$, consistent RMSE is $15\%$ below matched AA in this
single seeded scene/noise realization. The dashed two-capture reference
is $\sqrt2\sigma/\operatorname{std}(I_{1/2})=0.20$ on the same central
crop; it is not an exact prediction-noise floor, which depends on
operator row norms.

Composition compares $\Kop{2/3}\Kop{1/2}I_1$ with
$\Kop{1/3}I_1$: relative errors are $3.8\times10^{-5}$, $0.154$
and $0.244$ for consistent, cubic AA and matched AA, respectively.
Each AA arm uses its own direct result; its second step follows the
same dilation rule at $2/3$ in intermediate pixels. Shared input noise
need not cancel under different operators, although no independent
target noise enters. The consistent residual includes numerical and
cutoff approximations. Thirteen telescope factors recover the camera
within maximum multiplier error $2\times10^{-4}$ on $|\omega|\le0.85\pi$.

The Fourier arm's $3.10\times10^{-3}$ finite-grid composition residual
reflects a fine period ($2^{15}$ points) indivisible by strides 24,
48, and 72, plus interpolation of a discontinuous cutoff. On a
compatible 1,440-sample periodic grid, direct Fourier evaluation
closes within $9.5\times10^{-14}$ maximum absolute error. The main
arm results retain their original boundary realization.

The spectral ratio compares $I_{1/2}$ at frequency $\nu$ with
$I_1$ at $\nu/2$. Welch estimates use Hann windows of 256 and
512 samples respectively, half overlap, and squared Fourier
magnitudes normalized by the squared window sum.
The measured ratio follows
$|\hK{0}(\nu)/\hK{0}(\nu/2)|^2$ until noise dominates near $0.7$
of coarse Nyquist. It differs from cubic AA's squared transfer and
Fourier's unit response below cutoff. The released script reproduces
these rounded values.

{\section{Clean one-shot upscaling (Table~\ref{tab:clean-upscale})}\label{app:clean-upscale}
We use all five Set5 RGB images at $r\in\{2,3,4\}$. ResizeRight
generates floating-point observations with reflected boundaries,
centered grids, and output lengths $\lceil N/r\rceil$. Separate
degradations are Keys bicubic AA and a Gaussian with standard deviation
$2$ HR pixels, truncated at $4\sigma$ and normalized per phase.
There is no noise, quantization, divisibility crop, or target-selected
regularizer; all methods share the input and restore original HR size.

ResizeRight supplies bilinear, bicubic, and Lanczos-3 interpolation;
we also test ideal Fourier interpolation without camera equalization.
Camera-derived refinement uses the Fourier multiplier
$H(\omega/r)/H(\omega)$ on source-visible frequencies, applied
separably, followed by evaluation on the same centered output grid.
For the Gaussian anchor, camera variance is $4/(r^2-1)$, giving
multiplier $\exp(2\omega^2/r^2)$ per dimension. For the bicubic-AA
anchor, writing $U$ for the continuous Keys transfer function gives
$H(\omega)=\prod_{j\ge0}U(\omega/r^j)$ within the digital band;
the refinement multiplier telescopes to $1/U(\omega)$.
There is no extra taper, threshold, or ridge. Cosine quadrature tabulates
the Keys transform at 8193 frequencies from zero to Nyquist.

Fourier realization uses each input's complete even-reflection period
(length $2n-2$). Nyquist terms are split before the fractional shift
to ResizeRight's grid. With output
length $N$ and input length $n$, that grid is
$x_j=j/r+(n-1)/2-(N-1)/(2r)$.
Analytic cosine checks, including Nyquist modes and nondivisible
sizes, agree within $1.3\times10^{-14}$; repeating the reflection
period agrees to numerical precision.

Finite sampled degradations still entail aliasing, truncation, and
boundaries, so do not satisfy the continuous camera model exactly.

We average per-image BT.601 Y-PSNR with peak one and a prespecified
$r$-pixel HR border shave. Primary scores clip predictions to $[0,1]$;
raw scores follow. Camera-derived refinement beats bicubic and
Lanczos-3 in all 30 Set5 image--degradation--scale combinations.
The comparison is with fixed interpolation, not other kernel-aware SR
or optimal natural-image reconstruction.

\begin{table*}[t]
\centering\small
\begin{tabular}{llrrrrr}
\hline
Degradation & Scale & Bilinear & Bicubic & Lanczos-3 & Fourier interp.
& Camera-derived\\
\hline
Gaussian & $\times2$ & 27.36 & 27.78 & 27.90 & 27.88 & 32.60\\
& $\times3$ & 26.93 & 27.50 & 27.77 & 27.80 & 31.24\\
& $\times4$ & 26.28 & 27.04 & 27.39 & 27.49 & 29.00\\
Bicubic AA & $\times2$ & 32.26 & 33.69 & 34.33 & 34.54 & 34.77\\
& $\times3$ & 29.50 & 30.42 & 30.86 & 30.96 & 31.06\\
& $\times4$ & 27.56 & 28.43 & 28.82 & 28.92 & 29.01\\
\hline
\end{tabular}
\caption{{Full clean-upscaling comparison, mean Set5 Y-PSNR in dB.
Only camera-derived refinement uses the known degradation.
All scores use the same observation and border.}}
\label{tab:clean-upscale-full}
\end{table*}

Unclipped camera-derived scores are $32.33$, $31.21$, $28.95$ dB for
Gaussian and $34.74$, $31.02$, $28.96$ dB for bicubic AA. A 32-pixel
boundary shave changes Gaussian $\times2$ from $32.60$ to $34.92$ dB;
the primary result uses the stated $r$-pixel shave. Per-image scores
and timings accompany the experiment. The full-reflection Fourier
implementation is slower than local interpolation, so these results
do not establish bicubic-level runtime.

\paragraph{Quantized Set5 observations.}
We also clip shared observations to $[0,1]$ and round to 8-bit.
Quantization makes exact Gaussian $\times2$ refinement ill-conditioned:
mean PSNR is $10.70$ dB clipped ($3.63$ raw), versus bicubic's
$27.78$ dB, through amplification of quantization error.
We apply \eqref{eq:wiener} using the two-dimensional separable camera
symbol, with $\eta=\sigma_q/\operatorname{std}(I)$,
$\sigma_q=1/(255\sqrt{12})$, and preserve the observed channel means.
This fixed uniform-quantization heuristic uses no target pixels and
is not necessarily optimal. It yields $32.59$ dB clipped ($32.55$ raw)
in that condition. The full comparison is:
\begin{center}\small
\begin{tabular}{llrrrr}
\toprule
Input blur & Up & Bicubic & Lanczos-3 & Exact camera & Regularized camera\\
\midrule
Gaussian & $\times2$ & 27.78 & 27.90 & 10.70 & 32.59\\
& $\times3$ & 27.50 & 27.77 & 30.90 & 30.95\\
& $\times4$ & 27.04 & 27.39 & 28.98 & 28.98\\
Bicubic AA & $\times2$ & 33.69 & 34.33 & 34.75 & 34.75\\
& $\times3$ & 30.42 & 30.85 & 31.05 & 31.05\\
& $\times4$ & 28.43 & 28.82 & 29.00 & 29.00\\
\bottomrule
\end{tabular}
\end{center}
The clean table isolates unquantized refinement; quantized or noisy
use may require regularization.

\paragraph{BSD100 and Set14.}
The same floating-point protocol covers all 100 BSD100 and 14 Set14
images. Tables~\ref{tab:clean-upscale}, \ref{tab:clean-upscale-full},
and~\ref{tab:clean-upscale-set14} report BSD100, Set5, and Set14.
Camera-derived refinement leads each dataset mean in all six
conditions, without a per-image claim.

\begin{table*}[t]
\centering\small
\begin{tabular}{llrrrrr}
\hline
Degradation & Scale & Bilinear & Bicubic & Lanczos-3 & Fourier interp.
& Camera-derived\\
\hline
Gaussian & $\times2$ & 25.46 & 25.76 & 25.86 & 25.85 & 29.81\\
& $\times3$ & 25.11 & 25.53 & 25.72 & 25.77 & 28.26\\
& $\times4$ & 24.64 & 25.16 & 25.39 & 25.43 & 26.34\\
Bicubic AA & $\times2$ & 29.34 & 30.37 & 30.82 & 30.93 & 31.06\\
& $\times3$ & 27.00 & 27.65 & 27.95 & 28.03 & 28.12\\
& $\times4$ & 25.54 & 26.11 & 26.33 & 26.34 & 26.36\\
\hline
\end{tabular}
\caption{{Full clean-upscaling comparison, mean Set14 Y-PSNR in dB.
Only camera-derived refinement uses the known degradation.
Same floating-point protocol as Table~\ref{tab:clean-upscale-full}.}}
\label{tab:clean-upscale-set14}
\end{table*}}

{\section{Shared-solver kernel transfer}
\label{app:transfer}
Both arms receive the same exact $\times2$ relative kernel and predict
$\times3$ and $\times4$ kernels with the same finite forward operator,
grid, boundaries, solver, and regularization. No HR target selects
these choices. This synthetic known-anchor test isolates the scale
law; it is neither blind SR nor another RealSR experiment.

\paragraph{Independent optical targets.}
We use central $192\times192$ RGB crops of all five Set5 images as
periodic scenes. The two separable cameras have one-dimensional
symbols (i) $H(\omega)=\exp[-(4/3)\omega^2/2]$, a Gaussian
idealization whose $\times2$ blur has standard deviation 2 HR pixels,
and (ii) the autocorrelation of
$P(u)=\exp[2\pi i(0.18u^2+0.10u^4)]\mathbf1_{[-1,1]}(u)$,
computed with 8193 pupil samples, cutoff $0.9\pi$, and full-pitch
pixel integration. For a scene spectrum $F$, the HR reference is
rendered with $H(\omega_y)H(\omega_x)$; the factor-$S$ LR capture
is independently rendered with $H(S\omega_y)H(S\omega_x)$ and read
at coordinates $Sj+(S-1)/2$. We test $S=2,3,4$ clean and with
white RGB noise of standard deviation $1/255$ (three paired draws).
The target is an HR camera capture, unlike the unfiltered reference
in Table~\ref{tab:clean-upscale}.

\paragraph{Identical reconstruction, different transferred kernels.}
The supplied anchor is $U_2(\omega)=H(2\omega)/H(\omega)$ on the
visible band. The consistent arm reconstructs $H$ using 40 telescope
factors and uses $H(S\omega)/H(\omega)$; the stretch arm uses
$U_2(S\omega/2)$. Both equal the supplied anchor at $S=2$.
Each arm's one-dimensional matrix $A$ applies its multiplier on the
same centered periodic grid, so
$D(X)=AXA^\top$. Both solve
\[
\min_X \|D(X)-Y\|_F^2+\eta^2\|X\|_F^2
\]
on the mean-subtracted channels by separable SVD, then restore the
observed channel means. Primary $\eta=0.01$ and sensitivity levels
$0.001,0.003,0.03,0.1$ were fixed before scoring. Maximum absolute
telescope, independent forward-rendering, and adjoint discrepancies
are $1.37\times10^{-13}$, $7.69\times10^{-14}$, and $1.25\times10^{-14}$.

\paragraph{Scope of the Gaussian realization.}
The Gaussian arm samples a cutoff-free ratio, including weak output
aliases in its finite forward matrix. It is an approximation;
unprojected Gaussian sampling does not meet the alias-free identities.
The pupil cutoff is below output Nyquist.

For Y-PSNR we clip, shave $S$ HR border pixels, average noise draws
within images, then average image dB scores. Paired $95\%$ intervals
bootstrap five images 10,000 times (seed 200927), without treating
draws as independent scenes. At $S=2$, even noisy outputs coincide.
The primary transfer results are:
\begin{center}\small
\begin{tabular}{llrrrrl}
\toprule
& & \multicolumn{2}{c}{Clean} & \multicolumn{2}{c}{Noise $1/255$} & Noisy paired gain\\
Camera & Up & Stretch & Consistent & Stretch & Consistent & Mean [95\% CI]\\
\midrule
Gaussian & $\times3$ & 36.62 & 38.33 & 31.50 & 32.63 & $1.12\ [0.90,1.37]$\\
Gaussian & $\times4$ & 31.75 & 33.09 & 30.02 & 31.31 & $1.29\ [0.94,1.54]$\\
Pupil & $\times3$ & 37.62 & 43.28 & 34.10 & 34.53 & $0.43\ [-0.84,1.80]$\\
Pupil & $\times4$ & 32.68 & 38.23 & 31.36 & 33.77 & $2.41\ [0.95,3.88]$\\
\bottomrule
\end{tabular}
\end{center}
With a common anchor and solver, transfer favors the consistent law
in these means, but not uniformly under noise: two images favor
stretch for noisy pupil $\times3$, whose interval includes zero.

\paragraph{Regularization sensitivity.}
The table gives consistent-minus-stretch mean gains (dB). Under noise,
weak regularization can favor the more damped stretch arm. The primary
level was prespecified.
\begin{center}\small
\begin{tabular}{lllrrrrr}
\toprule
Camera & Up & Noise & $\eta=.001$ & $.003$ & $.01$ & $.03$ & $.1$\\
\midrule
Gaussian & $\times3$ & 0 & 4.73 & 3.25 & 1.71 & 0.87 & 0.24\\
& $\times3$ & $1/255$ & -0.74 & 0.35 & 1.12 & 0.87 & 0.25\\
& $\times4$ & 0 & 3.78 & 2.48 & 1.34 & 0.58 & 0.07\\
& $\times4$ & $1/255$ & -0.46 & 1.11 & 1.29 & 0.60 & 0.07\\
Pupil & $\times3$ & 0 & 5.74 & 5.74 & 5.66 & 4.82 & 1.01\\
& $\times3$ & $1/255$ & -3.95 & -2.16 & 0.43 & 2.88 & 1.01\\
& $\times4$ & 0 & 6.11 & 6.05 & 5.55 & 3.74 & 0.39\\
& $\times4$ & $1/255$ & -3.79 & -0.88 & 2.41 & 3.32 & 0.40\\
\bottomrule
\end{tabular}
\end{center}}

{\section{Sensitivity to estimated-kernel error}\label{app:psf-misspec}
We repeat the shared-solver test with unchanged captures and the same
smoothly perturbed $\times2$ anchor in both arms. For 20 unselected
random seeds, we use
$\widetilde U_2(\omega)=U_2(\omega)
[1+\alpha\sum_{j=1}^3c_j(\cos j\omega-1)]$, where the $c_j$ are
zero-mean Gaussian draws normalized by $\sum_j|c_j|$, giving smooth,
even, DC-preserving, sign-symmetric errors. We define the relative
$\times2$-anchor error by
\[
\varepsilon^2=
\frac{\sum_\omega |H(\omega)|^2
|\widetilde U_2(\omega)-U_2(\omega)|^2}
{\sum_\omega |H(\omega)|^2|U_2(\omega)|^2}
\]
and set it to $1\%$, $2\%$, or $5\%$ on the sampled Fourier grid. Grid, boundaries,
solver, $\eta=.01$, and three paired noise draws per image stay fixed.
No target or score selects a perturbation.

The table gives five-image mean camera-law minus AA-stretch PSNR (dB).
Noisy input has standard deviation $1/255$; $\pm$ is standard deviation
over perturbation seeds, and parentheses count AA-favoring seeds.
Zero error reproduces the preceding experiment; at $\times2$ the arms
tie for every perturbation.
\begin{center}\small
\begin{tabular}{llrrrr}
\toprule
Camera & $\varepsilon$ & Clean $\times3$ & Clean $\times4$
& Noisy $\times3$ & Noisy $\times4$\\
\midrule
Gaussian & $0\%$ & $+1.71$ & $+1.34$ & $+1.12$ & $+1.29$\\
& $1\%$ & $+1.70$ & $+1.35$ & $+1.12\pm0.13\ (0/20)$ & $+1.29\pm0.15\ (0/20)$\\
& $2\%$ & $+1.51$ & $+1.28$ & $+1.06\pm0.26\ (0/20)$ & $+1.24\pm0.31\ (0/20)$\\
& $5\%$ & $+0.37$ & $+0.68$ & $+0.66\pm0.70\ (8/20)$ & $+0.82\pm0.84\ (8/20)$\\
\midrule
Pupil & $0\%$ & $+5.66$ & $+5.55$ & $+0.43$ & $+2.41$\\
& $1\%$ & $+5.55$ & $+5.49$ & $+0.43\pm0.28\ (0/20)$ & $+2.40\pm0.22\ (0/20)$\\
& $2\%$ & $+5.14$ & $+5.26$ & $+0.39\pm0.57\ (8/20)$ & $+2.32\pm0.45\ (0/20)$\\
& $5\%$ & $+2.92$ & $+3.84$ & $+0.07\pm1.45\ (8/20)$ & $+1.77\pm1.26\ (1/20)$\\
\bottomrule
\end{tabular}
\end{center}
Small smooth errors preserve most gains, but some $2\%$ pupil
$\times3$ errors reverse the ranking. Magnitude alone does not
determine the result. Perturbed anchors need not be nonnegative
spatial PSFs; no particular estimator is modeled.

A separate coherent $10\%$ Gaussian-width bias gives $6.22\%$ weighted
anchor error and noisy $\times3/\times4$ gains of $-0.88/-0.88$ dB.
It nearly cancels AA stretching's Gaussian under-blur: width increases
of $8.9\%$ and $11.8\%$ match the true $\times3$ and $\times4$
multipliers. This is not typical random kernel-estimation noise.}

{\section{Optical prediction beyond alias-free sampling (Section~\ref{sec:limitations})}
\label{app:aliasing}
We vary optical cutoff over $\{0.9,1.5,2,3\}\pi$ at fixed sensor
pitch and pupil shape. The one-dimensional pupil is
$P(u)=\exp[2\pi i(0.18u^2+0.10u^4)]$, $|u|\le1$;
its normalized autocorrelation includes full-pitch pixel integration.
We render periodic continuous scenes on a grid 16 times finer than
the source sensor and independently sample aligned grids at factors
$1$, $1/2$, $1/3$, and $1/4$. Doubling fine-grid sampling changes
captures by less than $10^{-12}$ for the same trigonometric scene.

Two scene ensembles have $1/|\omega|$ amplitude (floored below
$2\pi/200$), or flat amplitude below $1.5$ radians and $1/|\omega|$
above it. Each has five fixed realizations, mean $0.5$, standard
deviation at most $0.12$, and range within $[0.02,0.98]$. We compare
clean and input-only white-noise ($\sigma=0.005$) cases against clean
captures. These one-dimensional signals are not an image benchmark.

For integer reduction $r$, the projected law applies
$H(r\omega)/H(\omega)$ where $|r\omega|<\pi$, then decimates. The
full-source-band heuristic omits the output-band restriction. Matched
AA equals the projected half-scale kernel and dilates it by $r/2$
elsewhere. Bicubic AA is an unmatched library-style reference. A
unit-sum discrete least-squares filter of radius $12r$ is fitted on
a separate scene four times longer than each test scene, with relative
SVD cutoff $10^{-8}$; test targets do not select coefficients.

We measure error against independently rendered captures and the
discrepancy between two half-steps and one quarter-step. Prediction
RMSE uses target standard deviation; path RMSE uses direct-output
standard deviation. Beyond sensor Nyquist, the projected family
still composes numerically but loses optical accuracy
(Figure~\ref{fig:aliasing-stress}). In the clean textured condition at
cutoff $3\pi$, its mean quarter-scale relative error is $20.86\%$,
versus $0.61\%$ for the full-band heuristic; the latter has $6.16\%$
path discrepancy. At cutoff $0.9\pi$, both predict the optical target
and compose to numerical precision.

The heuristic does not solve aliasing generally: frequencies $0.3\pi$
and $1.7\pi$ have identical cosine samples. Weighting their scene
amplitudes by opposite camera responses yields identical source but
different half-scale captures at cutoff $3\pi$. No operator of source
samples can recover both. Algebraic consistency and optical accuracy
therefore diverge outside the alias-free assumptions.

\begin{figure*}[t]
\centering
\includegraphics[width=0.7\textwidth]{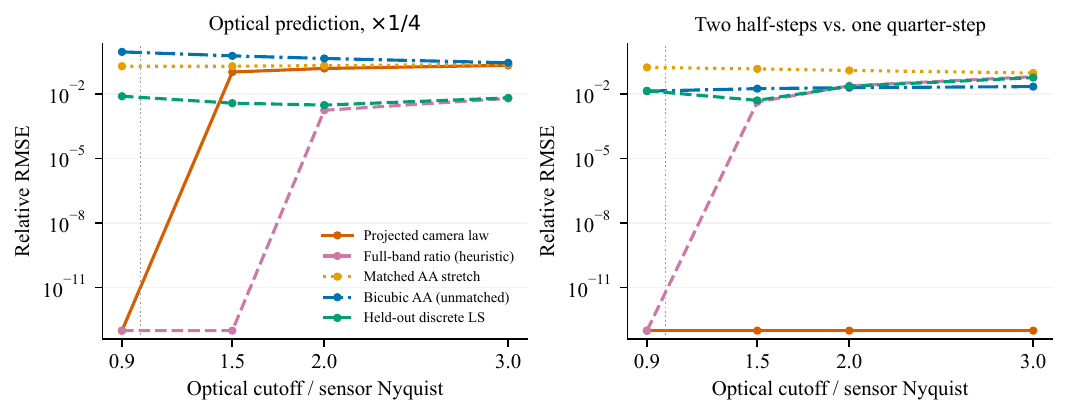}
\caption{{Aliasing stress test, clean textured ensemble, five scenes.
Left: prediction of an independently rendered quarter-scale capture.
Right: two half-steps versus one quarter-step. The projected camera
law remains algebraically consistent but ceases to reproduce aliased
optical captures. The full-band ratio predicts these captures more
accurately here, but is only a heuristic and loses composition.
The vertical line marks sensor Nyquist; values below $10^{-13}$ are
displayed at that numerical floor.}}
\label{fig:aliasing-stress}
\end{figure*}}

\section{Learned super-resolution
(Section~\ref{sec:applications}, Figure~\ref{fig:zssr})}
\label{app:zssr}

\paragraph{Task and controlled comparison.}
We test the effect of the degradation scale law on a learned inverse with
simplified, image-specific ZSSR~\cite{shocher18zssr}, not an optimized
reproduction of its full pipeline. Each BSD100, Set5~\cite{bevilacqua2012},
and Set14 image is reduced by $S\in\{2,3,4\}$. Within each kernel family and
scale, methods share the observation. For each image, method, and
$m\in\{1,\ldots,10\}$, a fresh network learns to reverse reduction by
$r=S^{1/m}$ and is then applied $m$ times. The network, training budget,
initialization, crop/augmentation schedule, and bicubic upsampling are
shared; only the training downsampler changes. No external training data,
back-projection, self-ensemble, injected noise, or augmentation from
intermediate reconstructions is used. Ground truth generates observations
and evaluates final outputs; it supplies no training targets and does not
select checkpoints or trigger early stopping.

\paragraph{Images and resampling.}
Original RGB images become floats in $[0,1]$ without cropping dimensions
to multiples of $S$. ResizeRight~\cite{resizeright} generates observations
at scale $1/S$ with reflection padding and the specified kernel; they are
neither clipped nor requantized (stored PNGs are previews only). Filters
are separable and identical across RGB. ResizeRight evaluates each
fractional kernel offset and normalizes weights at every output position.
Its extra anti-aliasing is disabled for custom kernels, which incorporate
their scale law; ordinary bicubic-AA uses its built-in cubic and AA stretching.

\paragraph{Band-limited bicubic family.}
Let $u$ be the Keys cubic with parameter $-1/2$, as implemented by
ResizeRight, and $U$ its continuous Fourier transform. Frequencies and
offsets use current-input pixels. Each total reduction $S$ has a formal
camera transfer function
\begin{equation}
H_S(\omega)=W(\omega)\prod_{j=0}^{17}U(\omega/S^j).
\label{eq:zssr-camera}
\end{equation}
The fixed $C^4$ window is $W(\omega)=1-p(t)$, where
\begin{align}
t&=\min\!\left(1,\max\!\left(0,
\frac{|\omega|/\pi-0.8}{0.2}\right)\right),\nonumber\\
p(t)&=126t^5-420t^6+540t^7-315t^8+70t^9.
\label{eq:zssr-window}
\end{align}
$W=1$ through $0.8\pi$ and vanishes at and beyond $\pi$.
The consistent and matched-AA multipliers are
\begin{equation}
Q_r(\omega)=\frac{H_S(r\omega)}{H_S(\omega)},
\qquad A_r(\omega)=U(r\omega)W(r\omega),
\label{eq:zssr-laws}
\end{equation}
on $|\omega|\le\pi/r$, and zero outside. At $r=S$, both use the same
anchor $U(S\omega)W(S\omega)$, avoiding finite-product discrepancies; it
also generates the observation. Input, anchor, output cutoff, and
sampling convention are shared; intermediate scale laws differ.
The inferred camera is a formal kernel model, not a claimed nonnegative
physical PSF.

We evaluate $U$ by trapezoidal cosine quadrature on 4097 spatial points
in $[0,2]$, tabulate it at 8193 frequencies in $[0,\pi]$, normalize its
DC value, and linearly interpolate. For $F_r\in\{Q_r,A_r\}$, the spatial
kernel is
\begin{equation}
k_r(x)=\frac{1}{\pi}\int_0^{\pi/r}
F_r(\omega)\cos(\omega x)\,d\omega.
\label{eq:zssr-kernel}
\end{equation}
The integral uses 4097 uniform frequencies. The even kernel is tabulated
at 512 offsets per pixel, linearly interpolated, and zero beyond
$R=32S$ (64, 96, or 128 input pixels). The roll-off controls spatial
tails; support-doubling checks on observed images passed an RMSE threshold
of $2\times10^{-5}$. Finite support, quadrature, and phase normalization
make the filters approximate rather than exactly band-limited, so they
need not compose to machine precision.

\paragraph{Gaussian family.}
The common observation kernel has standard deviation $\sigma=2$ in
original-image pixels for every $S$: the degradation width, not the
unstretched AA base width. At factor $r$, the current-input-pixel widths are
\begin{equation}
\sigma_{\rm ours}(r)=2\sqrt{\frac{r^2-1}{S^2-1}},
\qquad \sigma_{\rm AA}(r)=\frac{2r}{S}.
\label{eq:zssr-gaussian}
\end{equation}
The AA width dilates the shared full-scale anchor, with no intermediate
tuning. We evaluate the Gaussian density directly, truncate at
$|x|=4\sigma$, and use ResizeRight's per-position normalization, with no
spectral cutoff, taper, ridge, or minimum-width constraint. Both widths
equal two at $r=S$ and use the same kernel callable. The untruncated
continuous law composes exactly; truncation and sampling may not.
Fixed $\sigma=2$ suppresses aliasing less at $S=4$ than at $S=2$.

\paragraph{Sampling-grid control.}
Ours and AA share the ResizeRight grid. For input length $n$, output
length $n'$, and reduction $r$, the input-coordinate sample centers are
\begin{equation}
x_j^{\rm RR}=rj+\frac{n-1}{2}-\frac{r(n'-1)}{2},
\qquad n'=\lceil n/r\rceil.
\label{eq:zssr-rr-grid}
\end{equation}
AA (PyTorch grid) retains the AA kernel, normalization, and reflection,
but uses
\begin{equation}
x_j^{\rm grid}=r(j+\tfrac12)-\tfrac12,
\qquad n'=\lfloor n/r\rfloor.
\label{eq:zssr-grid-control}
\end{equation}
All upsampling remains ResizeRight-centered: only the training
downsampler's grid changes, not the end-to-end PyTorch resize, and this
arm does not isolate the scale law. It is trained independently at
$m=1$ because its sampled anchor may differ. For both main-figure
families, ours and matched AA share the one-step downsampler and trained
model; their $m=1$ values are not independent measurements.

\paragraph{Network and online pairs.}
The CNN has eight stride-one $3\times3$ convolutions, seven 64-channel
hidden layers with ReLU, and three input/output channels (225,091
parameters). Its residual connection adds the bicubically interpolated
input to the CNN output. Biases are zero and weights Kaiming-normal,
except the zero-initialized final convolution; thus every model starts
at bicubic interpolation.
There is no batch normalization or output activation.

Each update draws a fresh crop uniformly from the observation, applies
one of eight $90^\circ$ rotation/reflection combinations, downsamples it
by the selected method, then bicubically restores its dimensions. The
undegraded crop is the target. Pairs are generated \emph{after} cropping,
not from a precomputed downscaled image. Only crop coordinates,
augmentations, and resampling weights are cached; weights come from
ResizeRight's action on an identity basis and are checked against direct
resizing.

The supervised square has side
$c=\min(64,\min(h_{\rm LR},w_{\rm LR})-16)$.
On Set5, all $\times2$ crops have $c=64$; at $\times3$, only woman has
$c=60$. At $\times4$, baby, bird, butterfly, head, and woman have
$c=64,56,48,54,41$, respectively. Before degradation, crops include a
reflected $R+\lceil2r\rceil+10$-pixel halo per side; reflection extends
the observation, but the supervised crop stays within its boundaries.
After downsampling and interpolation, the CNN receives the central
$(c+16)\times(c+16)$ region; loss uses the central $c\times c$ output,
excluding convolution-padding effects.

\paragraph{Optimization and implementation.}
Each model uses 10,000 Adam updates, batch size one, RGB $\ell_1$ loss,
$\beta_1=0.9$, $\beta_2=0.999$, $\epsilon=10^{-8}$, and no weight decay.
The learning rate warms linearly to $10^{-4}$ in 20 updates, drops to
$10^{-5}$ after 6000 and $10^{-6}$ after 8500; the final checkpoint is
used. NumPy and PyTorch seeds reset to zero for each run, pairing
initialization, crops, and augmentations across methods. Runs use
PyTorch 2.8.0+cu129 on an
RTX 5090. Image operations use float32 and metrics float64; TF32 is
enabled for CNN convolutions but not resampling matrix products.
ResizeRight weights are built on CPU for $S=2$ and GPU for $S=3,4$,
consistently across arms. cuDNN benchmarking precludes guaranteed
bitwise reproducibility despite seeds. We follow original ZSSR commit
\texttt{d2948526} and ResizeRight commit \texttt{510d4d5b}.

\paragraph{Repeated inference.}
From the common observation, each of $m$ steps bicubically upsamples
by $r$ and applies the same residual CNN. Step $j$ has side length
$\lceil n_0S^{j/m}\rceil$, with $10^{-9}$ tolerance before rounding;
the last output takes the original dimensions. The requested factor
remains $r$ despite size rounding. Each CNN application has an
eight-pixel reflected border, removed afterward. Intermediate outputs
are neither clipped nor quantized; there is no stage-specific training,
registration, shift correction, or intensity fitting. The gray reference
is one-shot ResizeRight bicubic interpolation of the same observation
to the same final size, without a network.

\paragraph{Evaluation and interpretation.}
For plotted metrics, only final predictions are clipped to $[0,1]$.
We exclude $S$ high-resolution border pixels per side and compute
BT.601 Y-PSNR with peak one: the RGB error is weighted by
$(65.481,128.553,24.966)/255$ (the luminance offset cancels). Curves
average image-level dB scores, not pooled pixel errors. We retain
unclipped predictions, RGB and unclipped Y-PSNR, and output ranges.
All ten steps are shown with panel-specific vertical ranges. These
one-seed comparisons are neither significance tests nor standard
bicubic leaderboard scores.

Degradation consistency does not ensure improvement with $m$ because
learning, boundary, sampling, and finite-support errors remain. On Set5,
our Gaussian $S=4,m=10$ result falls below bicubic interpolation.
Late-step Gaussian AA losses include amplitude instability, not just blur:
at $S=2,m=10$, Set5 mean AA Y-PSNR is 7.80 dB clipped versus $-12.25$ dB
unclipped. These reflect iteration of a learned inverse, not one-shot
resizing gains; the cause of our own decline is not uniquely identified.
We re-scored saved predictions and checked checkpoints for finite values
and full budgets; crop-coordinate hashes verify pairing. Earlier completed
results were retained wholesale, regardless of PSNR.

\begin{figure*}[t]
\centering
\includegraphics[width=0.8\textwidth]{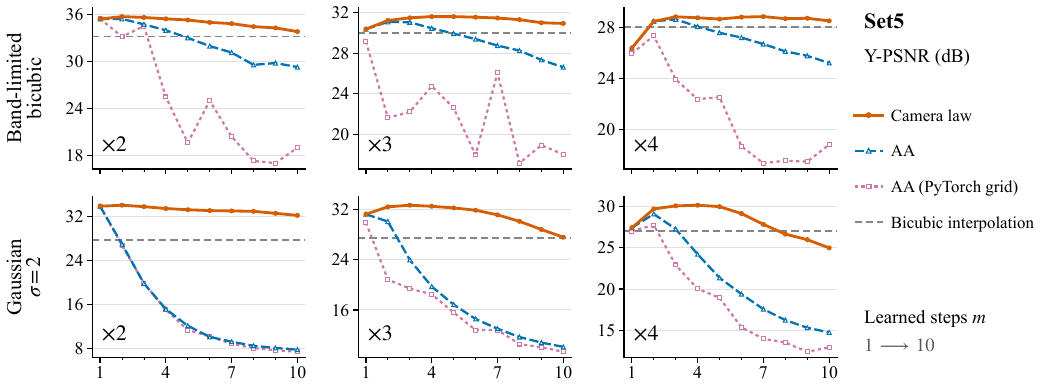}
\caption{{\footnotesize Set5 repeated learned SR, corresponding to
Fig.~\ref{fig:zssr}. Mean Y-PSNR over five images, one seed; gray is
one-shot bicubic. Vertical ranges differ. Protocol: App.~\ref{app:zssr}.}}
\label{fig:zssr-set5}
\end{figure*}

\begin{figure*}[t]
\centering
\includegraphics[width=0.8\textwidth]{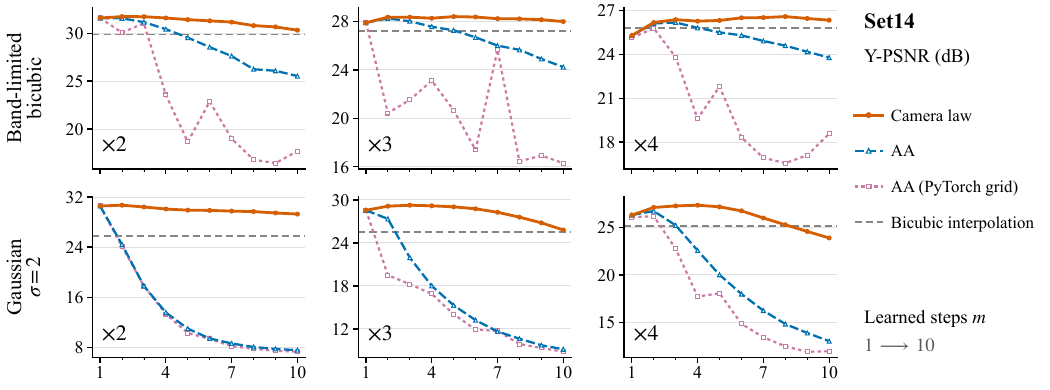}
\caption{{\footnotesize Set14 repeated learned SR, corresponding to
Fig.~\ref{fig:zssr}. Mean Y-PSNR over 14 images, one seed; gray is
one-shot bicubic. Vertical ranges differ. Protocol: App.~\ref{app:zssr}.}}
\label{fig:zssr-set14}
\end{figure*}

\paragraph{Cross-dataset curves.}\label{app:zssr-datasets}
Figure~\ref{fig:zssr} covers all 100 BSD100 images;
Figs.~\ref{fig:zssr-set5} and~\ref{fig:zssr-set14} cover five Set5 and
14 Set14 images. All favor the camera law over matched AA at late
steps, but not over one-shot bicubic at every step.

\begin{figure*}[h]
\centering
\includegraphics[width=0.8\textwidth]{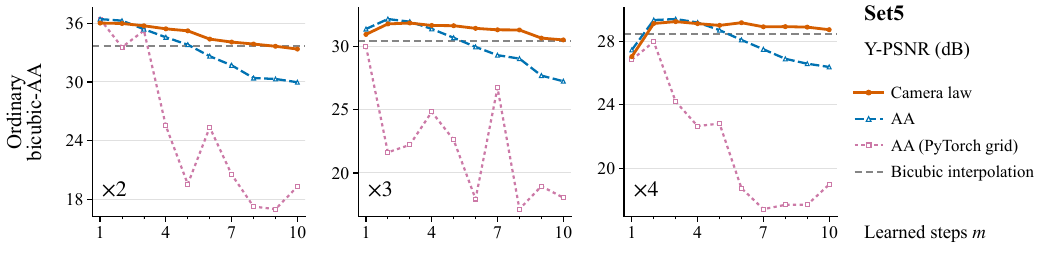}
\caption{Learned SR from ordinary bicubic-AA observations. Same
Set5 protocol described above, without band-limiting the
observation kernel. Ours uses the untapered camera ratio on the
retained band, so its anchor is approximate: the one-step results
need not coincide. AA wins at early steps; ours is more robust over
longer chains, but does not attain the best mean PSNR over the sweep.
Curves average all five images, one seed; gray is one-shot bicubic
interpolation. AA (PyTorch grid) changes only downsampling grids,
not the shared ResizeRight upsampler. Vertical ranges differ.}
\label{fig:zssr-bicubic}
\end{figure*}

{\paragraph{Best step count versus late-step robustness.}
We also report the largest mean PSNR in the ten-step sweep, with its
step count in parentheses. These descriptive maxima use the same five
Set5 images, not held-out step-count selection or new test scores.
\begin{center}\small
\begin{tabular}{llrrrr}
\toprule
Observation family & Up & Ours (m) & AA (m) & AA grid (m) & Bicubic interp.\\
\midrule
Band-limited bicubic & $\times2$ & 35.74 (2) & 35.44 (2) & 35.46 (1) & 33.18 \\
Band-limited bicubic & $\times3$ & 31.61 (4) & 31.08 (2) & 29.17 (1) & 29.96 \\
Band-limited bicubic & $\times4$ & 28.83 (7) & 28.62 (3) & 27.33 (2) & 28.01 \\
Gaussian & $\times2$ & 34.12 (2) & 33.94 (1) & 33.92 (1) & 27.78 \\
Gaussian & $\times3$ & 32.70 (3) & 31.27 (1) & 29.94 (1) & 27.50 \\
Gaussian & $\times4$ & 30.17 (4) & 29.11 (2) & 27.77 (2) & 27.04 \\
Ordinary bicubic AA & $\times2$ & 36.01 (1) & 36.41 (1) & 36.42 (1) & 33.69 \\
Ordinary bicubic AA & $\times3$ & 31.84 (3) & 32.16 (2) & 30.02 (1) & 30.42 \\
Ordinary bicubic AA & $\times4$ & 29.23 (3) & 29.39 (3) & 27.96 (2) & 28.43 \\
\bottomrule
\end{tabular}
\end{center}
Best-mean margins over AA are much smaller than ten-step margins;
for ordinary bicubic observations, AA has the higher maximum at every
factor. The many-step result concerns robustness of the learned inverse
to its training scale law, not whether ten steps is the best SR setting.}

{\paragraph{Additional paired training seeds.}
On Set5, we repeated the protocol with seeds 1 and 2 at $m=1,5,10$.
Balanced completed conditions cover band-limited bicubic at all three
factors and Gaussian at $\times2$; Gaussian $\times3,\times4$ repeats
remain incomplete. Each reported condition retains all five images
and both arms. As before, the one-step arms share a trained model.
We checked source and input hashes, full 10,000-update budgets, and
paired crop-coordinate hashes. Completed runs were neither selected
nor discarded by PSNR.

The table combines seed 0 with seeds 1 and 2. Intervals resample five
image-level gains averaged over seeds (10,000 draws, seed 200928);
the last column is the sample standard deviation of three seed-specific
mean gains. Seeds are not treated as 15 independent images.
Figure~\ref{fig:zssr-set5} retains the original ten-step sweep.
\begin{center}\small
\begin{tabular}{llrrrrr}
\toprule
Family & Up & $m$ & Ours & AA & Paired gain [95\% CI] & Seed SD\\
\midrule
Band-limited bicubic & $\times2$ & 1 & 35.44 & 35.44 & 0.00 $[0.00,0.00]$ & 0.000 \\
Band-limited bicubic & $\times2$ & 5 & 35.30 & 33.01 & 2.29 $[1.75,2.94]$ & 0.013 \\
Band-limited bicubic & $\times2$ & 10 & 33.84 & 29.21 & 4.63 $[3.84,5.44]$ & 0.086 \\
Band-limited bicubic & $\times3$ & 1 & 30.36 & 30.36 & 0.00 $[0.00,0.00]$ & 0.000 \\
Band-limited bicubic & $\times3$ & 5 & 31.58 & 29.95 & 1.64 $[1.37,1.91]$ & 0.026 \\
Band-limited bicubic & $\times3$ & 10 & 30.80 & 26.57 & 4.24 $[3.84,4.66]$ & 0.075 \\
Band-limited bicubic & $\times4$ & 1 & 26.34 & 26.34 & 0.00 $[0.00,0.00]$ & 0.000 \\
Band-limited bicubic & $\times4$ & 5 & 28.68 & 27.53 & 1.14 $[0.90,1.33]$ & 0.098 \\
Band-limited bicubic & $\times4$ & 10 & 28.51 & 25.30 & 3.21 $[2.46,3.80]$ & 0.070 \\
Gaussian & $\times2$ & 1 & 33.93 & 33.93 & 0.00 $[0.00,0.00]$ & 0.000 \\
Gaussian & $\times2$ & 5 & 33.29 & 12.17 & 21.13 $[17.86,23.44]$ & 0.184 \\
Gaussian & $\times2$ & 10 & 32.30 & 7.64 & 24.66 $[21.46,27.29]$ & 0.203 \\
\bottomrule
\end{tabular}
\end{center}
Completed repeats corroborate late-step separation on these five scenes;
the intervals describe uncertainty, not broad dataset generalization.}

\paragraph{Ordinary bicubic observations.}
Figure~\ref{fig:zssr-bicubic} uses observations from unmodified
ResizeRight bicubic-AA; AA and the grid arm use its stretched cubic.
For ours, we omit $W$ from \eqref{eq:zssr-camera}, use the same 18-factor
product, and evaluate its ratio with \eqref{eq:zssr-kernel}, the same
quadrature, support, and interpolation. There is no smooth taper or
ridge, but the hard boundary $|\omega|\le\pi/r$ remains; the quotient
is not full-spectrum. Ordinary bicubic has out-of-band energy, so
agreement on the retained band does not ensure an equal sampled anchor.
Ours is trained independently at $m=1$, averaging about 0.4 dB below
AA at each $S$. On Set5 it overtakes AA from $m=3,4,5$ for $S=2,3,4$,
respectively; ten-step gains are 3.39, 3.26, and 2.35 dB. AA still
attains a higher best mean at every scale. The result supports greater
many-step robustness, not universal superiority or exact sampled
consistency under aliasing.

\section{Additional figures (Sections~\ref{sec:prelim}--\ref{sec:audit})}\label{app:figs}

\begin{figure*}[t]
\centering
\includegraphics[width=0.78\textwidth]{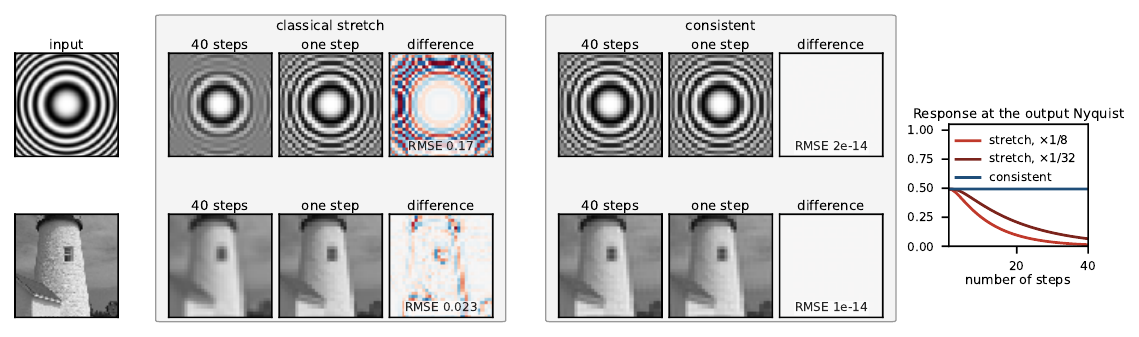}
\caption{\footnotesize The chain test: $\times\frac18$ in forty small steps
against one step, on a photograph and a zone plate. Right: detail kept
at the output Nyquist against the number of steps.}
\label{fig:chain}
\medskip
\includegraphics[width=0.24\textwidth]{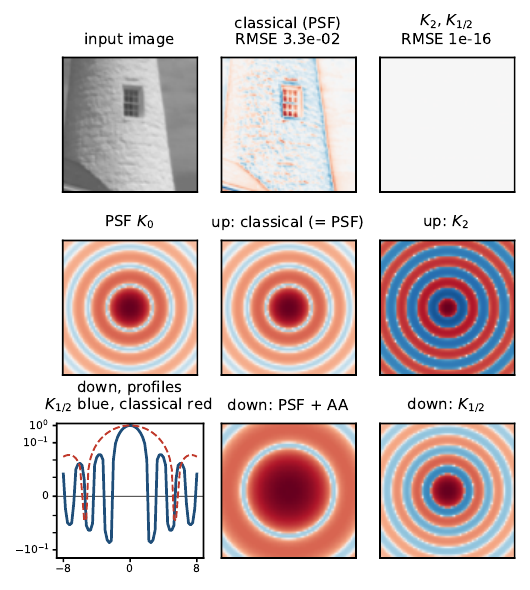}\hfill
\includegraphics[width=0.24\textwidth]{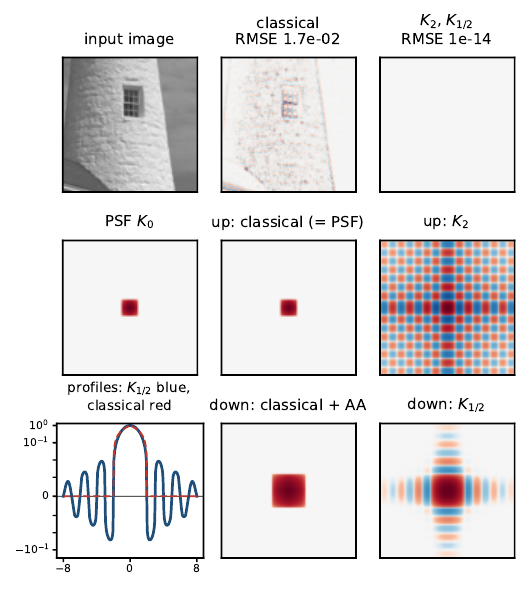}\hfill
\includegraphics[width=0.24\textwidth]{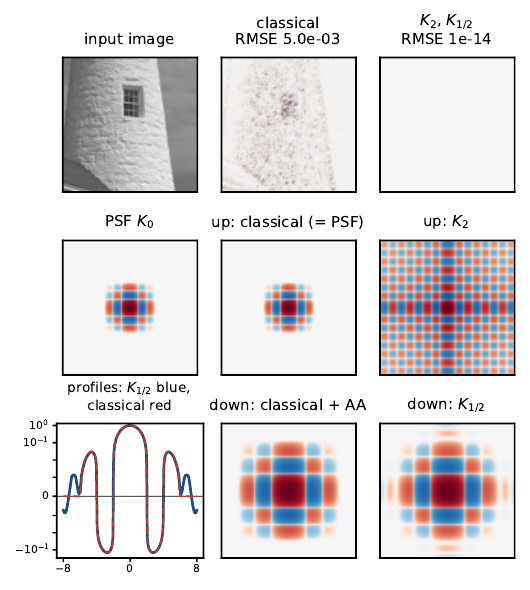}
\caption{\footnotesize Cycle test of Figure~\ref{fig:consistency}(a):
diffraction-limited camera, bilinear, and Lanczos-3 models. Tested
same-kernel baselines fail the round trip; consistent operators close
it to numerical precision.}
\label{fig:cycleall}
\end{figure*}

\begin{figure*}[t]
\centering
\begin{minipage}[t]{0.48\textwidth}
\centering
\includegraphics[width=\linewidth]{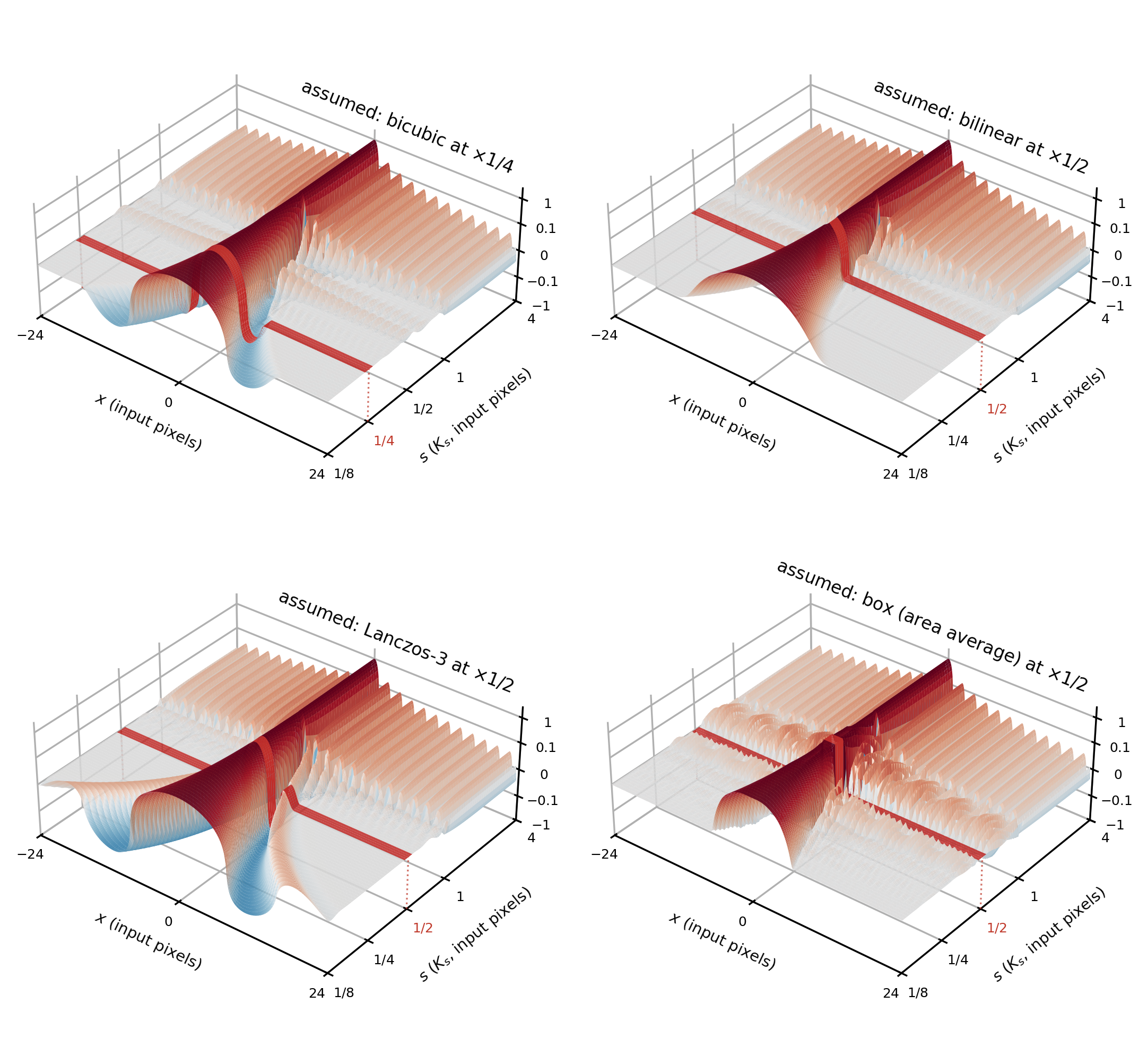}
\caption{\footnotesize Families from one anchor (red band): bicubic at
$\times\frac14$; bilinear, Lanczos-3, and box at $\times\frac12$.
Slices are peak-normalized, with signed-log heights. Telescoping gives
the camera endpoint, which changes if the bicubic anchor scale changes
(Remark~\ref{rem:aa}). Sinc rings persist; the tent rounds and the box
becomes bell-shaped. Ratios can diverge near source zeros (the canonical
multiplier is zero at exact zeros). The $3\%$ denominator cutoff makes
these regularized illustrations, not exact identities.}
\label{fig:surfslices}
\end{minipage}\hfill
\begin{minipage}[t]{0.48\textwidth}
\centering
\includegraphics[width=\linewidth]{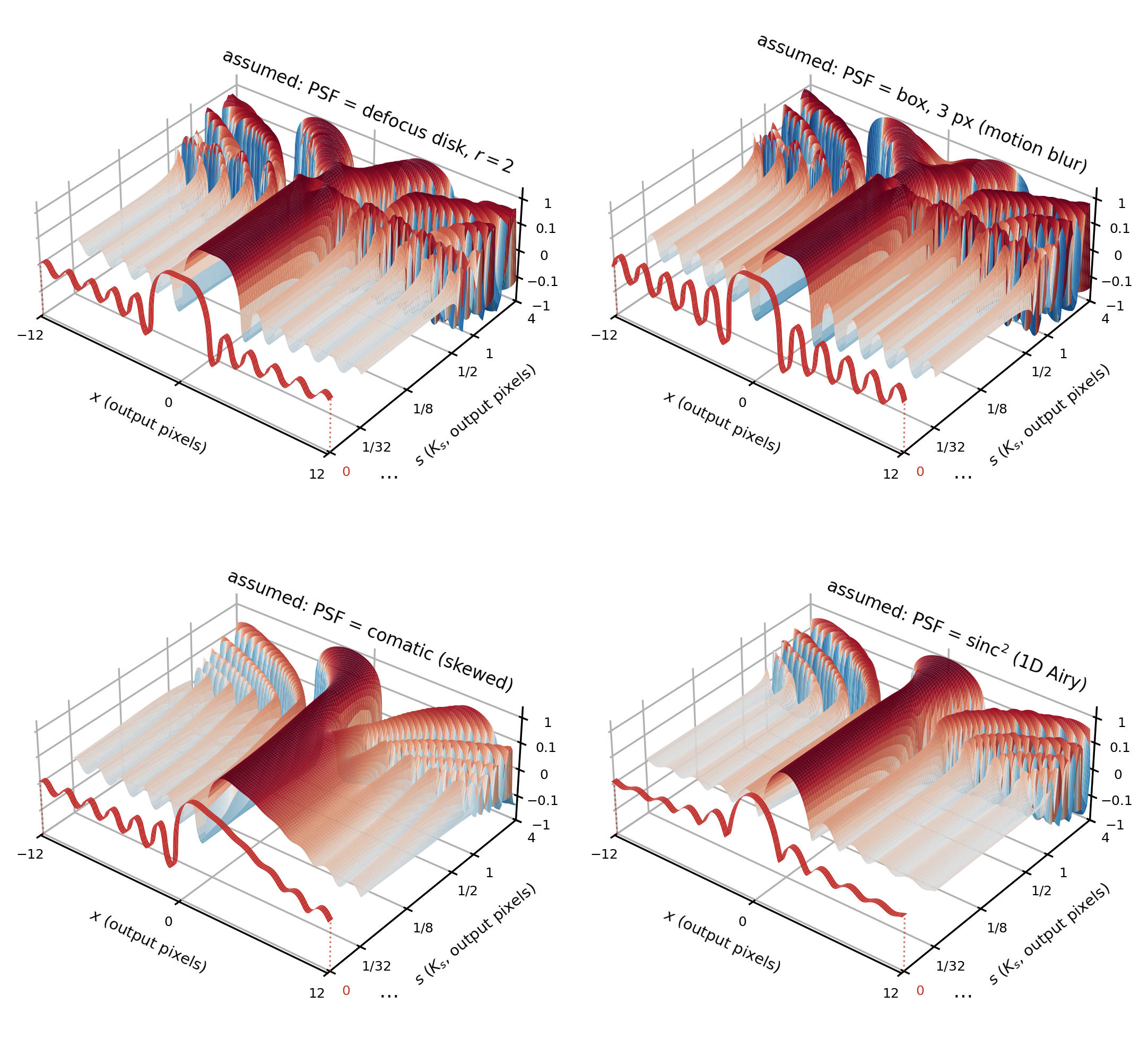}
\caption{\footnotesize Formal endpoints in output pixels (red bands
beyond the axis break); slices are peak-normalized with signed-log
heights. The first three models truncate a radius-$2$ defocus disk,
width-$3$ motion box, and asymmetric coma-like profile at Nyquist:
alias-free sampling does not ensure a nonnegative physical PSF. The
fourth is a physical slit-aperture $\operatorname{sinc}^2$ PSF, with
triangular OTF cut at $0.9\pi$. Interior zeros cause shape changes
and ringing; asymmetry exceeds symmetric conventions. Ratios use the
same denominator cutoff as Figure~\ref{fig:surfslices}.}
\label{fig:surfpsfs}
\end{minipage}
\end{figure*}

\begin{figure*}[t]
\centering
\begin{minipage}[t]{0.48\textwidth}
\centering
\includegraphics[width=0.80\linewidth]{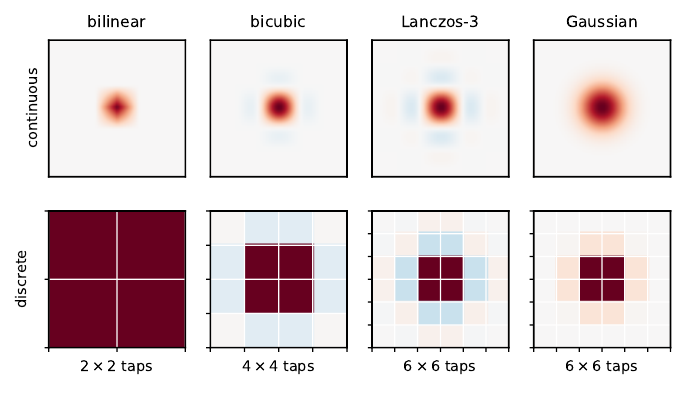}
\caption{\footnotesize Bilinear, bicubic~\cite{keys81}, Lanczos-3~\cite{duchon79},
and Gaussian kernels: continuous functions (top) and half-pixel taps
(bottom). Negative lobes survive as negative sampled values.}
\label{fig:prelim}
\end{minipage}\hfill
\begin{minipage}[t]{0.48\textwidth}
\centering
\includegraphics[width=0.72\linewidth]{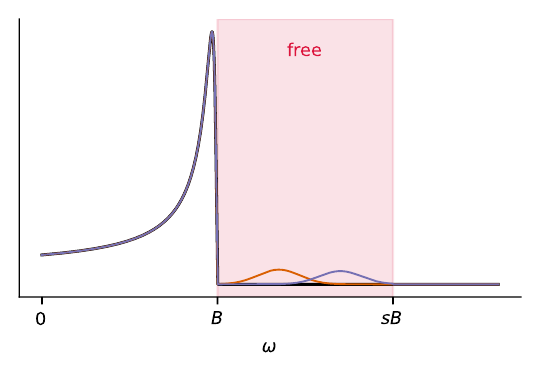}
\caption{\footnotesize Kernel freedom at $s=2$: black is forced on the
source band; colored bumps lie between source and target bands. Exact
$\Set{2}$ members require $Z$ to vanish on the source band and every
replica. Shown target-band freedom vanishes for $s\le1$, though the full
null space need not.}
\label{fig:set}
\end{minipage}

\medskip
\begin{minipage}[t]{0.48\textwidth}
\centering
\includegraphics[width=0.83\linewidth]{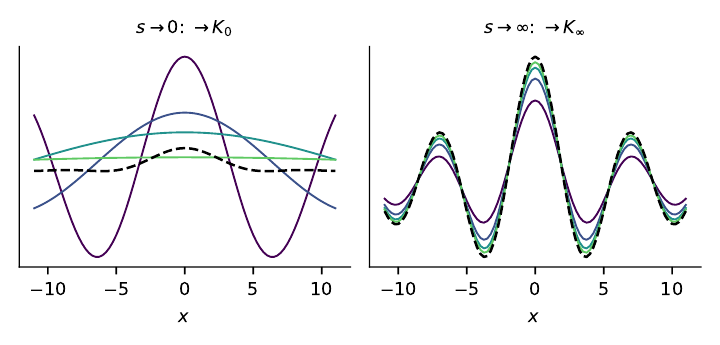}
\caption{\footnotesize Proposition~\ref{prop:endpoints} in kernels:
coarsening kernels in output coordinates converge to the PSF (left);
refining kernels converge to the hard-thresholded pseudo-inverse
(right). Dashed curves mark both limits.}
\label{fig:endpoints}
\end{minipage}\hfill
\begin{minipage}[t]{0.48\textwidth}
\centering
\includegraphics[width=0.96\linewidth]{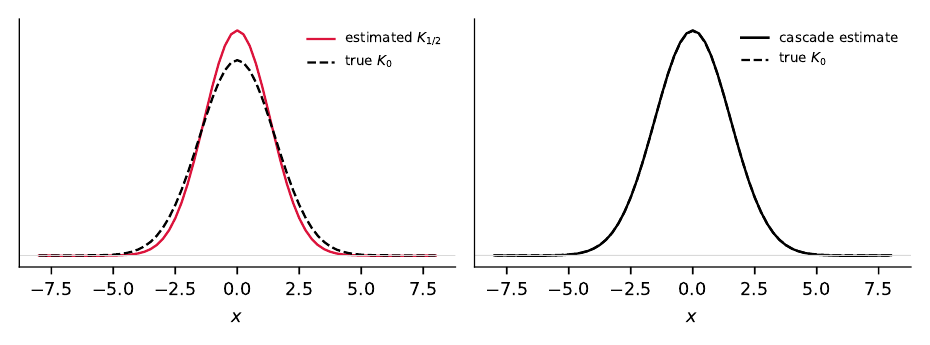}
\caption{\footnotesize Camera from one blind half-scale kernel
(Corollary~\ref{cor:telescope}). An estimated kernel with $3\%$
simulated error (left) yields about $1\%$ camera error through the
telescoping product (right). Route disagreement diagnoses
inconsistency, not a universal error bound.}
\label{fig:cascade}
\end{minipage}
\end{figure*}



\clearpage
\begin{thebibliography}{99}

\bibitem{burt83} P.~J. Burt and E.~H. Adelson. The Laplacian pyramid as a compact image code. \emph{IEEE Trans.\ Communications}, 31(4):532--540, 1983.

\bibitem{irani91} M.~Irani and S.~Peleg. Improving resolution by image registration. \emph{CVGIP: Graphical Models and Image Processing}, 53(3):231--239, 1991.

\bibitem{farsiu04} S.~Farsiu, M.~D. Robinson, M.~Elad, and P.~Milanfar. Fast and robust multiframe super resolution. \emph{IEEE Trans.\ Image Processing}, 13(10):1327--1344, 2004.

\bibitem{freeman02} W.~T. Freeman, T.~R. Jones, and E.~C. Pasztor. Example-based super-resolution. \emph{IEEE Computer Graphics and Applications}, 22(2):56--65, 2002.

\bibitem{dong16} C.~Dong, C.~C. Loy, K.~He, and X.~Tang. Image super-resolution using deep convolutional networks. \emph{IEEE Trans.\ PAMI}, 38(2):295--307, 2016.

\bibitem{keys81} R.~G. Keys. Cubic convolution interpolation for digital image processing. \emph{IEEE TASSP}, 29(6):1153--1160, 1981.

\bibitem{mitchell88} D.~P. Mitchell and A.~N. Netravali. Reconstruction filters in computer graphics. \emph{SIGGRAPH}, 1988.

\bibitem{duchon79} C.~E. Duchon. Lanczos filtering in one and two dimensions. \emph{J.\ Applied Meteorology}, 18:1016--1022, 1979.

\bibitem{talebi2021} H.~Talebi and P.~Milanfar. Learning to resize images for computer vision tasks. \emph{ICCV}, 2021.

\bibitem{iijima62} T.~Iijima. Basic theory on normalization of pattern (in case of typical one-dimensional pattern). \emph{Bulletin of the Electrotechnical Laboratory}, 26:368--388, 1962. In Japanese.

\bibitem{weickert99} J.~Weickert, S.~Ishikawa, and A.~Imiya. Linear scale-space has first been proposed in Japan. \emph{J.\ Math.\ Imaging Vis.}, 10(3):237--252, 1999.

\bibitem{witkin83} A.~P. Witkin. Scale-space filtering. \emph{IJCAI}, 1983.

\bibitem{koenderink84} J.~J. Koenderink. The structure of images. \emph{Biological Cybernetics}, 50:363--370, 1984.

\bibitem{lindeberg94} T.~Lindeberg. \emph{Scale-Space Theory in Computer Vision}. Kluwer, 1994.

\bibitem{lindeberg90} T.~Lindeberg. Scale-space for discrete signals. \emph{IEEE TPAMI}, 12(3):234--254, 1990.

\bibitem{lindeberg2024} T.~Lindeberg. Discrete approximations of Gaussian smoothing and Gaussian derivatives. \emph{J.\ Math.\ Imaging and Vision}, 66(5):759--800, 2024.

\bibitem{lindeberg2011} T.~Lindeberg. Generalized Gaussian scale-space axiomatics comprising linear scale-space, affine scale-space and spatio-temporal scale-space. \emph{J.\ Math.\ Imaging and Vision}, 40(1):36--81, 2011.

\bibitem{lindebergbretzner03} T.~Lindeberg and L.~Bretzner. Real-time scale selection in hybrid multi-scale representations. \emph{Scale-Space'03}, Springer LNCS 2695:148--163, 2003.

\bibitem{morel2011}{J.-M.~Morel and G.~Yu.
Is SIFT scale invariant?
\emph{Inverse Problems and Imaging}, 5(1):115--136, 2011.}

\bibitem{reyotero2016}{I.~Rey-Otero, J.-M.~Morel, and M.~Delbracio.
An analysis of the factors affecting keypoint stability in scale-space.
\emph{J.\ Math.\ Imaging Vis.}, 56(3):554--572, 2016.}

\bibitem{unser94} M.~Unser and A.~Aldroubi. A general sampling theory for nonideal acquisition devices. \emph{IEEE Trans.\ Signal Processing}, 42(11):2915--2925, 1994.

\bibitem{eldar2003} Y.~C. Eldar. Sampling with arbitrary sampling and reconstruction spaces and oblique dual frame vectors. \emph{J.\ Fourier Anal.\ Appl.}, 9(1):77--96, 2003.

\bibitem{huang2006}
B.~Huang and E.~M.-K. Lai.
Non-bandlimited resampling of images.
\emph{ICME}, pp.~149--152, 2006.

\bibitem{huang2009}
B.~Huang, E.~M.-K. Lai, and A.~P. Vinod.
Image resizing and rotation based on the consistent resampling theory.
\emph{ISPACS}, pp.~1--4, 2009.

\bibitem{unser1995resize}{M.~Unser, A.~Aldroubi, and M.~Eden.
Enlargement or reduction of digital images with minimum loss of information.
\emph{IEEE Trans.\ Image Processing}, 4(3):247--258, 1995.}

\bibitem{munoz2001}{A.~Mu\~noz, T.~Blu, and M.~Unser.
Least-squares image resizing using finite differences.
\emph{IEEE Trans.\ Image Processing}, 10(9):1365--1378, 2001.}

\bibitem{nehab2014} D.~Nehab and H.~Hoppe. A fresh look at generalized sampling. \emph{Foundations and Trends in Computer Graphics and Vision}, 8(1):1--84, 2014.

\bibitem{domke2009}{J.~Domke and Y.~Aloimonos.
Image transformations and blurring.
\emph{IEEE Trans.\ PAMI}, 31(5):811--823, 2009.}

\bibitem{cardoso2015} M.~J. Cardoso, M.~Modat, T.~Vercauteren, and S.~Ourselin. Scale factor point spread function matching: Beyond aliasing in image resampling. \emph{MICCAI}, 2015.

\bibitem{boucaud2016} A.~Boucaud, M.~Bocchio, A.~Abergel, F.~Orieux, H.~Dole, and M.~A. Hadj-Youcef. Convolution kernels for multi-wavelength imaging. \emph{Astronomy \& Astrophysics}, 596:A63, 2016.

\bibitem{delbracio2012}
M.~Delbracio, A.~Almansa, J.-M.~Morel, and P.~Mus\'{e}.
Subpixel point spread function estimation from two photographs at
different distances.
\emph{SIAM J.\ Imaging Sci.}, 5(4):1234--1260, 2012.

\bibitem{efrat2013} N.~Efrat, D.~Glasner, A.~Apartsin, B.~Nadler, and A.~Levin. Accurate blur models vs.\ image priors in single image super-resolution. \emph{ICCV}, 2013.

\bibitem{michaeli2013} T.~Michaeli and M.~Irani. Nonparametric blind super-resolution. \emph{ICCV}, 2013.

\bibitem{bellkligler19} S.~Bell-Kligler, A.~Shocher, and M.~Irani. Blind super-resolution kernel estimation using an internal-GAN. \emph{NeurIPS}, 2019.

\bibitem{shocher2020cc} A.~Shocher, B.~Feinstein, N.~Haim, and M.~Irani. From discrete to continuous convolution layers. \emph{arXiv:2006.11120}, 2020.

\bibitem{Huygens1690Traite} {C.~Huygens. \emph{Trait{\'e} de la lumi{\`e}re}. Pierre vander Aa, Leiden, 1690.}

\bibitem{goodman} J.~W. Goodman. \emph{Introduction to Fourier Optics}. Roberts \& Co., 3rd ed., 2005.

\bibitem{airy1835} G.~B. Airy. On the diffraction of an object-glass with circular aperture. \emph{Transactions of the Cambridge Philosophical Society}, 5:283--291, 1835.

\bibitem{resizeright} A.~Shocher. ResizeRight. GitHub repository, \url{https://github.com/assafshocher/ResizeRight}, 2018.

\bibitem{shannon49} C.~E. Shannon. Communication in the presence of noise. \emph{Proc.\ IRE}, 37(1):10--21, 1949.

\bibitem{nyquist28} H.~Nyquist. Certain topics in telegraph transmission theory. \emph{Trans.\ AIEE}, 47(2):617--644, 1928.

\bibitem{unser2000} M.~Unser. Sampling---50 years after Shannon. \emph{Proc.\ IEEE}, 88(4):569--587, 2000.

\bibitem{florack92} L.~M.~J. Florack, B.~M. ter~Haar~Romeny, J.~J. Koenderink, and M.~A. Viergever. Scale and the differential structure of images. \emph{Image and Vision Computing}, 10(6):376--388, 1992.

\bibitem{pauwels95} E.~J. Pauwels, L.~J. Van~Gool, P.~Fiddelaers, and T.~Moons. An extended class of scale-invariant and recursive scale space filters. \emph{IEEE TPAMI}, 17(7):691--701, 1995.

\bibitem{duits04} R.~Duits, L.~Florack, J.~de~Graaf, and B.~ter~Haar~Romeny. On the axioms of scale space theory. \emph{J.\ Math.\ Imaging Vis.}, 20:267--298, 2004.

\bibitem{moore1920} {E.~H. Moore. On the reciprocal of the general algebraic matrix. \emph{Bulletin of the American Mathematical Society}, 26:394--395, 1920. Abstract.}

\bibitem{penrose55} R.~Penrose. A generalized inverse for matrices. \emph{Proc.\ Cambridge Philos.\ Soc.}, 51:406--413, 1955.

\bibitem{benisrael03} A.~Ben-Israel and T.~N.~E. Greville. \emph{Generalized Inverses: Theory and Applications}. Springer, 2nd ed., 2003.

\bibitem{bernstein2014} G.~M. Bernstein and D.~Gruen. Resampling images in Fourier space. \emph{Publications of the Astronomical Society of the Pacific}, 126(937):287--295, 2014.

\bibitem{unser1993pyramid}{M.~Unser, A.~Aldroubi, and M.~Eden.
The $L_2$-polynomial spline pyramid.
\emph{IEEE Trans.\ Pattern Analysis and Machine Intelligence},
15(4):364--379, 1993.}

\bibitem{aldroubi2010invariance}{A.~Aldroubi, C.~Cabrelli,
C.~Heil, K.~Kornelson, and U.~Molter.
Invariance of a shift-invariant space.
\emph{J.\ Fourier Anal.\ Appl.}, 16(1):60--75, 2010.}

\bibitem{mallat89} S.~G. Mallat. A theory for multiresolution signal decomposition: The wavelet representation. \emph{IEEE Trans.\ Pattern Analysis and Machine Intelligence}, 11(7):674--693, 1989.

\bibitem{daubechies92} I.~Daubechies. \emph{Ten Lectures on Wavelets}. SIAM, 1992.

\bibitem{scherzer2001} O.~Scherzer and C.~Groetsch. Inverse scale space theory for inverse problems. \emph{Scale-Space 2001}, Springer LNCS 2106:317--325, 2001.

\bibitem{burger2006} M.~Burger, G.~Gilboa, S.~Osher, and J.~Xu. Nonlinear inverse scale space methods. \emph{Commun.\ Math.\ Sci.}, 4(1):179--212, 2006.

\bibitem{wanglee1998}{Y.-P.~Wang and S.~L.~Lee.
Scale-space derived from B-splines.
\emph{IEEE Trans.\ PAMI}, 20(10):1040--1055, 1998.}

\bibitem{lowe2004}{D.~G. Lowe. Distinctive image features from scale-invariant keypoints.
\emph{International Journal of Computer Vision}, 60(2):91--110, 2004.}

\bibitem{yu2011asift}{G.~Yu and J.-M.~Morel.
ASIFT: An algorithm for fully affine invariant comparison.
\emph{Image Processing On Line}, 1:11--38, 2011.}

\bibitem{worrall2019}{D.~Worrall and M.~Welling.
Deep scale-spaces: Equivariance over scale.
\emph{NeurIPS}, 32, 2019.}

\bibitem{lindeberg2026}{T.~Lindeberg.
Hybrid Lie semi-group and cascade structures for the generalized
Gaussian derivative model for visual receptive fields.
\emph{J.\ Math.\ Imaging Vis.}, 68:19, 2026.}

\bibitem{unser1995rotation}{M.~Unser, P.~Th\'evenaz, and L.~P. Yaroslavsky.
Convolution-based interpolation for fast, high-quality rotation of images.
\emph{IEEE Trans.\ Image Processing}, 4(10):1371--1381, 1995.}

\bibitem{scholkopf2002} B.~Sch\"olkopf and A.~J. Smola. \emph{Learning with Kernels}. MIT Press, 2002.

\bibitem{mezic2005} I.~Mezi\'c. Spectral properties of dynamical systems, model reduction and decompositions. \emph{Nonlinear Dynamics}, 41:309--325, 2005.

\bibitem{cai2019realsr} J.~Cai, H.~Zeng, H.~Yong, Z.~Cao, and L.~Zhang. Toward real-world single image super-resolution: A new benchmark and a new model. \emph{ICCV}, 2019.

\bibitem{opticallimits} OpticalLimits. Nikkor Z 24--70mm f/2.8 S review; Sony FE 24--70mm f/2.8 GM II review. \url{https://opticallimits.com}, 2026. Imatest MTF50 in LW/PH.

\bibitem{shocher18zssr} A.~Shocher, N.~Cohen, and M.~Irani. Zero-shot super-resolution using deep internal learning. \emph{CVPR}, 2018.

\bibitem{bevilacqua2012}
M.~Bevilacqua, A.~Roumy, C.~Guillemot, and M.-L.~Alberi Morel.
Low-complexity single-image super-resolution based on nonnegative
neighbor embedding. \emph{BMVC}, pages 135.1--135.10, 2012.

\bibitem{parmar2022} G.~Parmar, R.~Zhang, and J.-Y. Zhu. On aliased resizing and surprising subtleties in GAN evaluation. \emph{CVPR}, 2022.

\bibitem{hpatches_2017_cvpr}
V.~Balntas, K.~Lenc, A.~Vedaldi, and K.~Mikolajczyk.
HPatches: A benchmark and evaluation of handcrafted and learned local descriptors.
\emph{CVPR}, 2017.

\bibitem{rahman2023}{M.~A. Rahman and R.~A. Yeh.
Truly scale-equivariant deep nets with Fourier layers.
\emph{NeurIPS}, 36:6092--6104, 2023.}

\bibitem{zhang2019zoom}
X.~Zhang, Q.~Chen, R.~Ng, and V.~Koltun.
Zoom to learn, learn to zoom. \emph{CVPR}, pp.~3762--3770, 2019.

\end{thebibliography}
\end{document}